\documentclass[12pt]{article}
\usepackage{amssymb}
\usepackage{amsmath}
\usepackage{amsbsy}
\usepackage{graphicx}
\usepackage{enumerate}
\usepackage[authoryear]{natbib}
\usepackage{algorithm}
\usepackage{url} 
\usepackage{float}

\newcommand{\Y}{{\mbox{\boldmath $Y$}}}
\newcommand{\X}{{\mbox{\boldmath $X$}}}
\newcommand{\Q}{{\mbox{\boldmath $Q$}}}
\newcommand{\bP}{{\mbox{\boldmath $P$}}}
\newcommand{\mbf}[1]{\mbox{\boldmath${#1}$}}

\newtheorem{theorem}{Theorem}

\newcommand{\blind}{1}

\begin{document}

\def\spacingset#1{\renewcommand{\baselinestretch}%
{#1}\small\normalsize} \spacingset{1}

\title{False Discovery Rate Control for Lesion-Symptom Mapping with Heterogeneous data via Weighted P-values}

\if1\blind
{
  \title{\bf False Discovery Rate Control for Lesion Symptom Mapping with Heterogeneous data via Weighted P-values}
  \author{Siyu Zheng$^{a}$, Alexander C. McLain$^{a}$\thanks{\emph{email:} mclaina@mailbox.sc.edu} ,  Joshua Habiger$^{b}$, \\ 
  Christopher Rorden$^{c}$ and Julius Fridriksson$^{d}$ 
  \bigskip \\
$^a$Department of Epidemiology and Biostatistics,\\ University of South Carolina.\\
$^b$Department of Statistics, Oklahoma State University \\
$^c$Department of Psychology, University of South Carolina \\
$^d$Department of Communication Sciences and Disorders,\\ University of South Carolina\\
 } 
  \maketitle
  
} \fi

\if0\blind
{
  \bigskip
  \bigskip
  \bigskip
  \begin{center}
    {\LARGE\bf False Discovery Rate Control for Lesion-Symptom Mapping with Heterogeneous data via Weighted P-values}
\end{center}
  \medskip
} \fi

\begin{abstract}
Lesion-symptom mapping studies provide insight into what areas of the brain are involved in different aspects of cognition. This is commonly done via behavioral testing in patients with a naturally occurring brain injury or lesions (e.g., strokes or brain tumors). This results in high-dimensional observational data where lesion status (present/absent) is non-uniformly distributed with some voxels having lesions in very few (or no) subjects. In this situation, mass univariate hypothesis tests have severe power heterogeneity where many tests are known \textit{a priori} to have little to no power. Recent advancements in multiple testing methodologies allow researchers to weigh hypotheses according to side-information (e.g., information on power heterogeneity). In this paper, we propose the use of p-value weighting for voxel-based lesion-symptom mapping (VLSM) studies. The weights are created using the distribution of lesion status and spatial information to estimate different non-null prior probabilities for each hypothesis test through some common approaches. We provide a \emph{monotone minimum weight} criterion which requires minimum \emph{a priori} power information. Our methods are demonstrated on dependent simulated data and an aphasia study investigating which regions of the brain are associated with the severity of language impairment among stroke survivors. The results demonstrate that the proposed methods have robust error control and can increase power. Further, we showcase how weights can be used to identify regions that are inconclusive due to lack of power.
\end{abstract}

\noindent%

\spacingset{1.5}

\section{Introduction}
\label{sec1}

Data arising from neuroscience studies  have considerable statistical issues including a large number of parameters, an unknown spatial dependence structure, and (commonly) low statistical power. Neuroimaging consists of using magnetic resonance imaging (MRI), positron emission tomography (PET), electroencephalography (EEG), or other imaging modalities, to measure various aspects of brain structure and activity.  Data modalities from MRI include functional MRI (fMRI), structural T1 weighted images (T1), and diffusion-weighted imaging (DWI) among others.  These data are typically measured on a voxel level in three-dimensional space.  As imaging technologies improve the number of data voxels per scan has increased, possibly reaching into the millions depending on the spatial resolution of the image. Independent statistical tests are often computed for each location. Therefore, as spatial resolution increases, the opportunity for making erroneous discoveries increases.  This requires some principled thresholding to control for global type I error rate at a level $\alpha$.  Common criteria on the global type I error rate include the familywise error rate (FWER) \citep{Tuk94,NicHol02} and the false discovery rate (FDR) \citep{Benjamini1995}.  

Recent statistical methodology has considered using prior information about the hypotheses to improve results through p-value weighting \citep{Gen06, Roeder2009, Pena2011, Habiger2017, Ignatiadis2021, Lihua2018, Xianyang20}, grouping similar hypotheses \citep{Cai2009, Hu2010,Ignatiadis2016}, or weighting global type I error rate criteria \citep{BenHoc97,BenCoh17,Basetal18}. P-value weighting is a procedure that uses prior information on hypotheses heterogeneity to improve the overall power.  This prior information  -- commonly referred to as \textit{side-information} -- can consist of results from previous studies on the most `promising' hypotheses \citep{LiBar17,LiBar19,Lihua2018}, covariate data indirectly related to the hypotheses \citep{Ignatiadis2021}, or information related to the heterogeneity in the power functions of the hypotheses \citep{Pena2011,Habiger2017}.  The goal of a p-value weighting procedure is to design the weights to maximize the expected number of discoveries while controlling the FWER or FDR.

Modern weighting methods commonly use regression-type approaches to incorporate the side-information into the multiple testing procedure. Commonly these methods use the conditional two-group model where the side-information impacts the probability a test is null and the non-null p-value distribution. For example, \cite{Lihua2018} proposed adaptive p-value thresholding (AdaPT) which adaptively estimates a Bayes optimal p-value rejection threshold. This is done through the use of the Expectation Maximization (EM) algorithm using a set of partially masked p-values. A similar approach referred to as covariate adaptive multiple testing (CAMT) by \cite{Xianyang20}, also uses the EM algorithm with their M step being expressed in terms of the ratio of alternative and null distributions which is modeled using the beta density. \cite{Ignatiadis2021} proposed Independent Hypothesis Weighting (IHW) which divides all tests into several independent folds. For each fold, the estimated weight function can be learned from the p-values and covariates in the remaining folds. Similar to the AdaPT, IHW estimates the null probability and non-null distribution based on a conditional two-group model via an EM algorithm. AdaPT and IHW have been shown to provide finite sample FDR control, while CAMT can provide asymptotic FDR control. \cite{Cai2021} proposed a locally adaptive weighting and screening (LAWS) method to deal with spatial multiple testing problems. The LAWS procedure estimates the weights by using the spatial structure through a kernel screening method and can control the FDR asymptotically. \cite{BocLee18} proposed an FDR control multiple testing method \citep[R package \texttt{swfdr}][]{swfdrpackage} where -- in the spirit of the \cite{Sto02} procedure -- the unknown null indicator is replaced with an indicator the p-value is lower than some threshold. The indicators are used to estimate their associations with the side-information.

In this paper, we expand the Weighted Adaptive Benjamini Hochberg (WABH) procedure proposed by \cite{Habiger2017} to incorporate heterogeneous non-null probabilities and effect sizes. Further, we demonstrate how our methods are flexible to specific statistical models and are tailored to perform well in low-power settings, which are common in our application to voxel-based lesion-symptom mapping (VLSM) analyses \citep{Batetal03,Roretal07}. Further, WABH is known to be robust to poor estimation of the parameters governing the impact of the side-information on the weights. No previous methods are available that are designed for low-power settings and are robust to misspecification of the weights. In general, this procedure can be applied to any situation where the p-values arise from mass univariate logistic regression. Below, we detail the statistical issues that arise within VLSM and the solutions provided by our testing procedure.

\subsection{Voxel-based lesion symptom mapping}\label{sec.VLSM}

A main goal of research in neuroscience is to identify and examine areas of the brain related to behavioral or cognitive functions.  A common method is to use subjects with a recent brain injury (e.g., from a traumatic brain injury, epilepsy, or stroke) to map some domain of cognition to specific regions of the brain. This can provide theoretical insights regarding brain function and can also inform clinical treatment. The most popular lesion-symptom mapping approach -- voxel-based lesion-symptom mapping (VLSM) \citep{Batetal03,Roretal07}-- typically relies on structural MRI images (e.g., fMRI, T1, or DWI) where lesion status is measured on  parcellated three-dimensional voxels (e.g. 1 $mm^3$) and relates lesion status to an outcome of interest in each voxel \citep[see][for a recent review of the field]{Karetal17}.  The number of tests in a VLSM can reach millions depending on the resolution of the brain scan.

VLSM analyses are typically mass-univariate tests consisting of computing a simple test statistic (e.g., t-tests, General Linear Models, etc.) independently for each voxel and then using some multiple testing correction to identify significant associations with regions of the brain. There are a number of statistical issues that complicate such analyses.  First, since studies in humans cannot be designed to injure certain areas of the brain we must rely on naturally occurring injuries \citep{RorKar04}.  This commonly results in lesions being unequally distributed, with some areas/voxels having lesions in a few subjects. For example, stroke-related brain injury is determined by vasculature leading to some regions being far more vulnerable than others. Therefore, the spatial sampling of lesions is not random, and statistical power varies across space. Consider a study of language impairment following left hemisphere injury: we will have low power in regions typically spared in stroke and no power in the right hemisphere (as we have no variability). In response, some have advocated only using voxels that are impacted in, for example, 10\% of subjects to account for this issue \citep{Holetal96}.  Second, it is likely that the areas of the brain that will impact the cognitive outcome will have some spatial clustering. That is, the 3-dimensional coordinates of the voxels will be related to the non-null probability. Ignoring this relationship misses out on an important source of variation in the signal, thereby decreasing the overall power of the procedure.

The above issues naturally fit into the purview of weighted multiple testing. The naturally occurring injuries in VLSM create heterogeneous power among the voxels. Voxel power is a function of effect size, which is commonly unknown in practice.  In this paper, we propose and provide a straightforward solution consisting of a \emph{monotone minimum weight} criterion, which automatically estimates voxel power and has desirable properties for studies with low to moderate power. Further, we test using plug-in estimates of the non-null probability using state-of-the-art methods (AdaPT and CAMT) which utilize any spatial clustering of signals to gain power. In our data analysis, we demonstrate how the presentation of the impact of weighting is key for transparent reporting of weighted analyses.

The outline of the paper is as follows.  In Section 2, we review multiple testing procedures for data with heterogeneity among the hypotheses.  In Section 3, we discuss how weighted multiple testing procedures can be applied to VLSM in a number of common scenarios. In Section 4, we present results from numerous simulation studies that compare the performance of the proposed method to some common approaches. In Section 5, we present an analysis of $220$ individuals with chronic left hemisphere stroke and identify areas of the brain associated with the severity of aphasia, a language disorder that impacts the expression and comprehension of speech.

\section{Multiple Testing with Heterogeneous Data}\label{sec.MT_procs}
\subsection{Setup and Notation}
Consider testing null hypothesis $H_m$ based on the random vector $\mathcal{D}_m$ for $m = 1, 2, \ldots, M$. The decision to reject or retain $H_m$ with $\mbf{\mathcal{D}} = (\mathcal{D}_m; m = 1,2,...,M)$ is denoted by $\delta_m(\mbf{\mathcal{D}})\in\{0,1\}$ or $\delta_m$ for short, where $\delta_m$ is 1 if $H_m$ is rejected and is 0 otherwise.  For ease of exposition, we denote the event that a null hypothesis is true (false) by $H_m = 0$ ($H_m = 1$). Table \ref{BHtable} contains our notation for the total number of rejected and retained null hypotheses, incorrectly rejected and retained null hypotheses, correctly rejected and retained null hypotheses, and number of true and false null hypotheses.

The objective of most multiple testing procedures is to define the decision functions $\mathbf{\delta} = (\delta_m; m = 1, 2, ..., M)$ so as to maximize the expected number of true discoveries/positives $ETP = E[S]$, or minimize some type II error rate, such as the false non-discovery rate $FNR = {E[U]/E[M-R]}$ \citep{Sun2007}, subject to the constraint that the family-wise error rate $FWER = \Pr(V>0)$ or false discovery rate $FDR = E\left[V/R|R>0\right]\Pr(R>0)$ is not more than a pre-specified level $\alpha$. The FDR, or a variation of it such as the $mFDR = E[V]/E[R]$, is commonly utilized in large-scale multiple-hypothesis testing.  

\begin{table}[b]\center
\caption{\label{BHtable} Notation for various hypothesis testings subgroups based on if the null hypothesis is true ($H_m = 0$) or false ($H_m = 1$), and if the tests were rejected ($\delta_m = 1$) or not rejected ($\delta_m = 0$).}
\begin{tabular}{c|cc|c} \hline\hline
& $\delta_m = 0$ & $\delta_m = 1$ & Total \\ \hline
$H_m = 0$ & $T$ & $V$ & $M_0$ \\
$H_m = 1$ & $U$ & $S$ & $M_1$ \\ \hline
&$M - R$ & $R$ & M \\
\end{tabular}
\end{table}

\subsection{Weighted BH Methods}
Many multiple testing procedures have been developed for p-value statistics $\bP = (P_m; m = 1, 2,\ldots, M)$.  The basic idea is to find a p-value threshold $t$ for rejections and define $\delta_m(\bP) = I(P_m \leq t)$ where $I(\cdot)$ is the indicator function.  The well-known \cite{Benjamini1995}, or BH, procedure is implemented by finding the threshold $t_{BH} = \alpha k/M$ where $k = \max\left\{m:P_{(m)}\leq \alpha m/M\right\}$ and $P_{(m)}$ is the $m$th order p-value.  The BH procedure is then given by $\delta_m(\bP) = I(P_m \leq t_{BH})$. \cite{Benjamini1995} showed that if p-values from true null hypotheses are mutually independent and independent of p-values from false null hypotheses then this procedure has $FDR = \pi_0 \alpha \leq \alpha$, where $\pi_0 = M_0/M$ is the proportion of true null hypotheses. Adaptive FDR procedures (called ABH henceforth), leverage the fact that the BH procedure has FDR $=\pi_0\alpha$ by estimating $\pi_0$ via $\hat{\pi}_0$ and apply the BH procedure at level $\alpha/\hat\pi_0$ \citep{Storey2004}.  For example, \cite{Storey2004} proposed estimating $\pi_0$ and showed that if p-values are independent the ABH controls the FDR and is less conservative than the BH procedure.

Recent work has further improved upon the BH and ABH procedures by incorporating heterogeneity through p-value weighting.  For example, letting $w_1, w_2, ..., w_M$ be weights satisfying $M^{-1}\sum_mw_m = 1$, the weighted BH procedure (WBH) in \cite{Roeder2009} operates by applying the BH procedure to the weighted p-values denoted by $Q_m = P_m/w_m$. \cite{Roeder2009} showed that the WBH procedure provides FDR control under a finite mixture model for the p-values considered in \cite{Genovese2002}, among others. Optimal weights for the WBH procedure can depend on heterogeneous prior probabilities for the states of the null hypotheses \citep{Roeder2009,Hu2010,LiBar17,Lihua2018,Xianyang20,LiBar19}, heterogeneous power functions \citep{Pena2011} or both \citep{Cai2009,Ignatiadis2016,Habiger2017, Ignatiadis2018}.

\cite{Habiger2017} proposed applying the adaptive BH procedure to weighted p-values.  The procedure, henceforth called weighted adaptive BH (WABH),  operates as follows: (i) compute weighted p-values via $Q_m = P_m/w_m$ with $\Q = (Q_m; m = 1, 2,\ldots, M)$, (ii) estimate $\hat\pi_0 = \{\sum_m I(Q_m\geq \kappa) + 1\}/\{M(1-\kappa)\}$, (iii) compute threshold $t_{WABH} = \min\left\{\alpha, k \alpha/(\hat\pi_0 M)\right\}$ where $k = \max\left\{m: Q_{(m)}\leq m \alpha/(\hat\pi_0 M)\right\}$, and (iv) compute $\delta_m(\Q) = I(Q_m\leq t_{WABH})$. \cite{Habiger2017} showed that for reasonably specified weights the WABH procedure controls the FDR asymptotically and has higher ETP than the WBH and ABH procedures. In particular, as long as the utilized weights are positively correlated with optimal weights the procedure still controls the FDR and is more powerful than unweighted procedures. This allows for a procedure that incorporates heterogeneity across tests in applications where the precise nature and degree of heterogeneity isn't well known, but may be estimated or reasonably specified. The first step in utilizing such a procedure is to specify optimal Oracle weights.

\subsection{Optimal Oracle Weights}\label{sec.opt.weight}
Optimal Oracle weights are allowed to depend on heterogeneous prior probabilities and/or power functions.  Suppose, for example, $P_m$ has null CDF $F_0(t) = \Pr(P_m\leq t|H_m=0)=t$ and alternative CDF $F_m(t) = \Pr(P_m\leq t|H_m=1)$.  Further let $p_m = \Pr(H_m = 1)$ be the prior probability that $H_m$ is non-null for $m = 1, 2, ..., M$.   Suppose, for the moment, the Oracle situation where the weighted p-value threshold $t$ along with $p_m$ and $F_m$ for each $m$ are known.  The weighted p-value decision rule can be written $\delta_m(Q_m) = I(Q_m \leq t) = I(P_m \leq w_mt) \equiv I(P_m \leq t_m)$. 

The calculation of the optimal weights reduces to maximizing the expected number of true positives, $ETP = \sum_mp_mF_m(t_m)$ subject to the constraint that $M^{-1}\sum_mt_m = t$. That is, the objective is to find
$$\max_{\{t_m: m=1,\ldots,M\}}\left\{\sum_m p_m F_m(t_m)\right\} \mbox{ such that } G(\mbf{t};\mbf{p},\alpha)=0,$$
where $G(\mbf{t};\mbf{p},\alpha) = (1-\alpha)\sum_m (1-p_m)t_m - \alpha\sum_m p_m F_m(t_m)$ which can be solved via Lagrange multipliers \citep{Habiger2017}. 

Assuming $P_m$ arises from a normally distributed test statistic, the power of a test of size $t$ is $F_m(t) = \bar{\Phi}\left\{\bar{\Phi}^{-1}(t) - g_m\right\}$ where $\bar\Phi(\cdot) = 1- \Phi(\cdot)$, $\Phi(\cdot)$ is a standard normal distribution function, and $g_m$ the \textit{effect size over the standard error} of test $m$ (defined in Section \ref{sec.pow.est}). The expression for $f_m(t) = \frac{d}{dt}F_m(t)$ is 
\begin{equation}\label{eq.f}
f_m(t_m) = \frac{\phi\{\bar\Phi^{-1}(t_m) - g_m)\}}{\phi\{\bar\Phi^{-1}(t_m)\}},
\end{equation}
where $\phi(t)$ is the standard normal density function and
$$\frac{d}{dt_m}G(\mbf{t};\mbf{p},\alpha) = G'(t_m,p_m,\alpha) = (1-p_m)(1-\alpha)-\alpha p_mf_m(t_m).$$ 
Setting $p_mf_m(t_m) - \lambda G'(t_m,p_m,\alpha)=0$ yields the following expression
$$f_m(t_m)=\frac{\lambda(1-p_m)(1-\alpha)}{p_m(1+\lambda\alpha)} = c_m(\lambda), $$
where the solution in terms of $t_m$ is
\begin{equation} \label{eq.t}
 t_m(\lambda) = \bar\Phi\left[0.5g_m + \log\{c_m(\lambda)\}g_m^{-1}\right].
\end{equation}
The Lagrange multiplier is found by solving
\begin{equation}\label{eq.lambd}
\sum_m (1-p_m) t_m(\lambda) - \alpha\left[\sum_m (1-p_m) t_m(\lambda) + \sum_m p_m F_m\{t_m(\lambda)\}\right] = 0
\end{equation}
for $\lambda$.  The weights are then given by $w_m = t_m(\hat \lambda)\{M^{-1}\sum_m t_m(\hat \lambda)\}^{-1}$ where $\hat{\lambda}$ is the solution to (\ref{eq.lambd}). Once the weights are calculated the WABH procedure can be implemented.

\subsection{Dependence between tests}\label{sec.dep}
\cite{Storey2003} showed that the ABH procedure provides asymptotic FDR control under a weak dependence structure for the p-values and \cite{Habiger2017} extended this result to weighted p-values.  Weak dependence occurs, for example, when (weighted) p-values are correlated within groups but independent across groups. This structure may be reasonable to our application of interest because p-values are likely to be correlated within regions or clusters of voxels, but are nearly independent across distant regions. As a result, we ignore the dependence between p-values in our proposed testing procedure. Our simulation studies in Section \ref{sec.sim} explore the impact of dependence on our proposed method and other procedures by generating spatially dependent locations for non-null tests (to varying degrees) with spatially dependent data. Those results suggest that the degree of dependence does not have an impact on the FDR of the proposed method.

\section{Estimation of weights in VLSM}\label{sec.w.VLSM}
VLSM is a procedure that measures the strength of the association between lesion status and a cognitive outcome, independently for each voxel \citep{Batetal03}.  Let $X_{im}$ denote a measure of whether brain voxel $m$ has a lesion for person $i$, $m=1,2,\ldots,M$ and $i=1,2,\ldots,n$. Let $Y_i$ denote the outcome of interest for person $i$, which we assume to be continuous. Further, let $X^+_{im} = h(\sum_{j \neq m} X_{ij})$ be a measure of the total lesion size excluding voxel $m$ for person $i$ for some function $h$.  Below we consider $X^+_{im} \in \mathbb{R}$, but incorporating multidimensional $X^+_{im}$ is straightforward. Since voxel damage can only be a detriment to the cognitive outcome we consider $H_m$ to be one-sided hypothesis tests, however, the methods are easily generalized to two-sided tests.  We focus on logistic regression since it can model total lesion size ($X^+_{im}$) as a nuisance confounder \citep{Karetal04,Arnetal18}.

The optimal oracle weights in Section \ref{sec.opt.weight} require the specification of $g_m$ and $p_m$ in equations (\ref{eq.f}) and (\ref{eq.lambd}). To estimate the $p_m$'s we use existing general methods (discussed in Section \ref{sec.prior.prob}). In Section \ref{sec.pow.est}, the heterogeneous $g_m$'s are calculated by utilizing heterogeneous standard error calculations and prior knowledge that the power of the tests are low. Our resulting WABH algorithm is outlined in Section \ref{sec.MT_procs}.

\subsection{Estimation of prior null probabilities using known methods}\label{sec.prior.prob}

Ideally, values for the prior null probabilities ($p_m$) can be based on previous studies and expert knowledge.  When this is not possible, $p_m$ can be estimated based on the unweighted p-values. While there are many approaches to estimating $p_m$ in such cases, we focus on the AdaPT and CAMT procedures \citep{Lihua2018,Xianyang20}, due to the ease of implementing them in statistical software. Let $\mbf{z}_m$ denote the so-called `side-information' hypothesized to have an impact on $p_m$ and/or $f_m$, the density of $P_m$ under the alternative. Both AdaPT and CAMT use the two-groups model where $P_m|\mbf{z}_m,H_m \sim (1-H_m)f_0 + H_mf_m$ where $f_0$ is the density of $P_m$ under the null and $p_m = \Pr(H_m=1|\mbf{z}_m)$. To implement AdaPT or CAMT, we need to specify a parametric form for the relationship between (a) $\mbf{z}_m$ and $p_m$, and (b) $\mbf{z}_m$ and $f_m$.  For both models, $\log\{p_m/(1-p_m)\}$ is modeled via components of $\mbf{z}_m$. For (b), in AdaPT we assume $f_m$ is a beta density, and for CAMT we assume the ratio $f_m/f_0$ is a beta density. Specifically, a beta$(k_m,1)$ where $\log(k_m)$ is modeled via components of $\mbf{z}_m$. See \cite{Lihua2018}, and \cite{Xianyang20} for details on their estimation procedures. After AdaPT or CAMT are implemented, the estimates of $p_m$ are extracted and used in our testing procedure.

The WABH procedure can be implemented in various specifications of (a) and (b) above. For our simulation study, $\mbf{z}_m$ consists of the $2\times 2$ grid coordinates of test $m$, denoted by $\mbf{z}_m^p$, and the predicted standard error ($S_m$) denoted by $\mbf{z}_m^f$. In the simulation study for both AdaPT and CAMT, $\mbf{z}_m^p$ and $\log\{p_m/(1-p_m)\}$ are related using a linear combination of $5$ degree natural cubic splines for each coordinate and their interaction. Further, $\log(k_m)$ was modeled using a $5$ degree natural cubic spline on $\mbf{z}_m^f$. See Section \ref{sec.data.anal} for the specification of these relationships in our real data analysis. 


\subsection{Estimation of effect sizes and MMW criterion}\label{sec.pow.est}

In this section, we develop expressions for estimating $g_m$, which will be used to estimate optimal weights in subsequent sections.  We consider a strictly continuous outcome $Y_i \in \mathbb{R}$ and a binary $X_{im}\in (0,1)$ lesion indicator, which is modeled as a function of $Y_i$ and $X_{im}^+$ with a logistic regression model.  

Let $\mbf{Y}=\{Y_{1},Y_{2},\ldots,Y_{n}\},$ $\mbf{X}_m^+=\{X^+_{1m}, X^+_{2m}, \ldots, X^+_{nm}\}$ and $\mbf{X}_m=\{X_{1m},X_{2m},\ldots,X_{nm}\}$ for $m=1,\ldots,M$.   We consider null hypotheses $H_m:\beta_{1m}=0$ and alternative $H_m:\beta_{1m}>0$ where
$$\mbox{logit}\{\Pr(X_{im}=1|Y_i,X_{im}^+)\} = \beta_{0m} + \beta_{1m}Y_i + \beta_{2m} X^+_{im},$$
where $\mbox{logit}(p)=\log\{p/(1-p)\}$.  The Wald test p-values are $P_{m} = \bar\Phi(\hat \beta_{1m}/\widehat{SE}_m)$, where $\widehat{SE}_m$ is the estimated standard error of $\hat\beta_{1m}$. Clearly, the power of a test will depend upon $\beta_{1m}$ and $SE_m$.  Using previous results \cite{VaeSko04}, the standard error of $\hat\beta_{1m}$  can be approximated via
\begin{equation}\label{eq.logreg.SE}
SE_m = \left[\frac{(1-R^2_{m})}{n s^2_Y\bar X_m(1-\bar X_m)} \right]^{1/2} + o_p(n^{-1/2})
\end{equation}
where $R^2_{m}$ is the coefficient of determination for regressing $\Y$ on $\X_m^+$, $\bar{X_m}=\sum_i(X_{im})/n$, $s_Y^2=\sum_i(Y_{i}-\bar Y)^2/(n-1)$. Then,
\begin{equation}\label{eq.logreg.T}
\frac{\beta_{1m}}{SE_m} = \frac{\eta_{m}}{S_{m}} +o_p(n^{-1/2})
\end{equation}
where $\eta_{m} = \beta_{1m}s_Y$  and $S_{m}=[(1-R^2_{m})/\{n\bar X_m(1-\bar X_m)\}]^{1/2}$.  Assuming normality of the test statistics, given $g_m = \eta_{m}/{S}_{m}$ the power of a test of size $t$ is $F_m(t) = \bar{\Phi}\left\{\bar{\Phi}^{-1}(t) - g_m\right\}$.  Since $S_m$ can be estimated \textit{a priori}, the heterogeneity in the power can be calculated given the effect size $\eta_m$. To specify $\eta_m$, we consider the case where prior information differentiating $\eta_m$ is unavailable and set $\eta_m=\eta$ for all $m$.

\begin{figure}[t]
\centering
\includegraphics[scale=0.55]{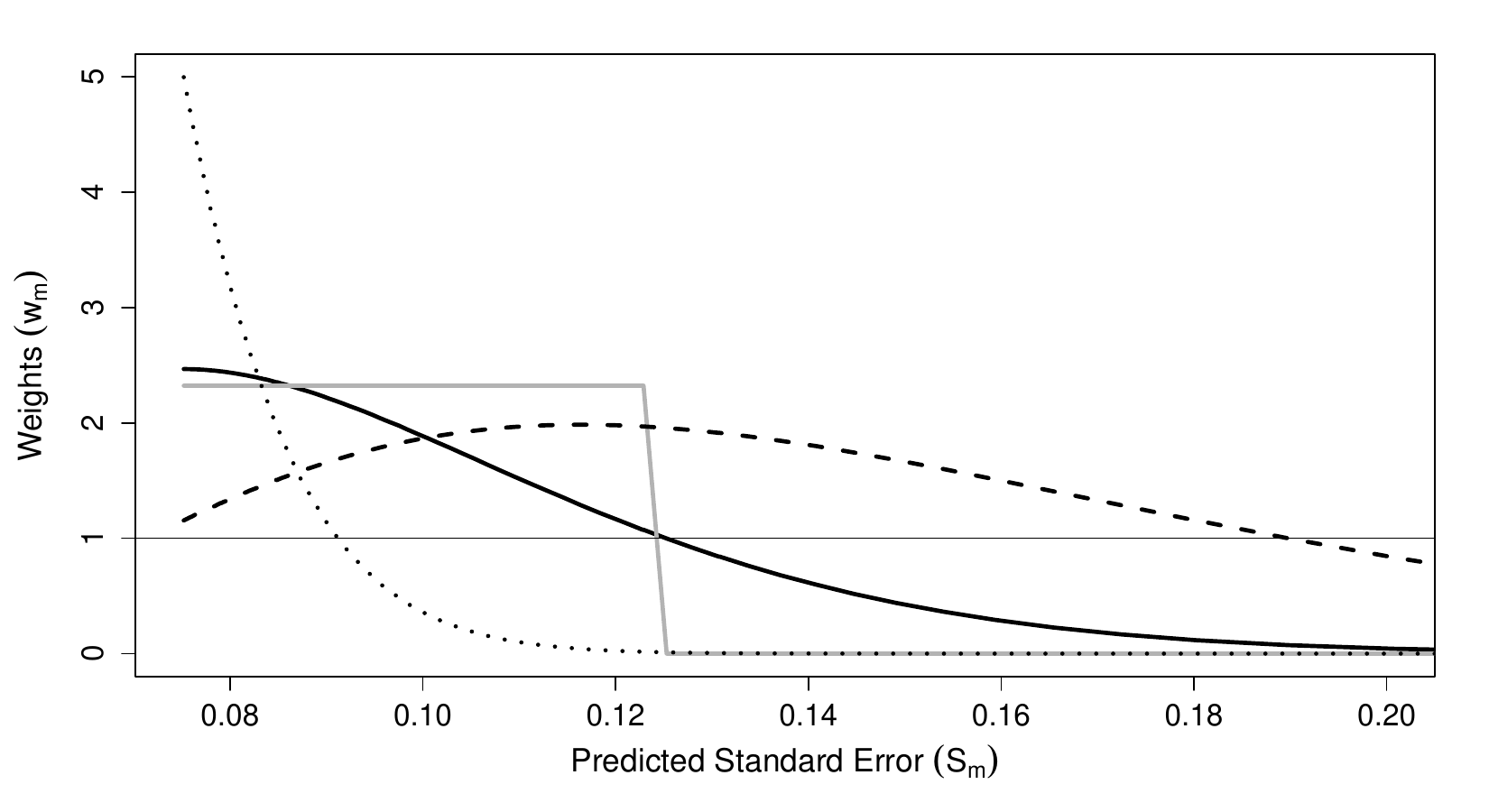}
\caption{\label{Fig.eta.pl} Estimated p-value weights ($w_m$) by their predicted standard error ($S_m$) for the aphasia data where $\eta = 0.1$ (black dotted), $0.178$ (black solid), and $0.25$ (black dashed), along with the 10\% rule (gray solid).}
\end{figure}

Figure \ref{Fig.eta.pl} displays the relationship between the weights ($w_m$) and $S_m$ for different $\eta$. This figure also shows the 10\% rule (which is common in practice), where the weights are given by $w_m = MI(\bar X_m \in [0.1,0.9])/(\sum_{k=1}^M I(\bar X_k \in [0.1,0.9]))$. Note that the 10\% rule up weights tests (i.e., $w_m \geq 1$) with high power (lower $S_m$) and down weight tests with low power.  That is, the weights are monotonically non-decreasing in power (or non-increasing in $S_m$). \cite{Habiger2017} showed that in low-power settings, optimal weighting schemes do involve increasing weight for tests with higher power and decreasing weights for those with lower power. Thus, the intuition behind the 10\% rule is correct, however, a more sophisticated weighting scheme from the optimal weights in Section \ref{sec.opt.weight} is available. We refer to our weights as the monotone minimum weights (MMW).  In summary, the MMW weights are the specific optimal weights that satisfy the desired monotonicity property while ensuring that weights are not too aggressive, i.e. no weight is too large (aggressive weighting schemes may result in very large weights for only a few tests and amount to not testing the vast majority of tests, which is intractable).  

Let us provide details. First, while $S_m$ in equation \eqref{eq.logreg.T} is a known source of heterogeneity affecting power, $\eta_m$ is not.  The MMW, for $p_m$'s computed as in the previous section, arises by choosing $\eta$ as large as possible so that weights are still monotone. The resulting weights are depicted in Figure \ref{Fig.eta.pl} by the thick black line. Note that in Figure \ref{Fig.eta.pl} the MMW weights amount to choosing $\eta = .178$.  Other values do not satisfy the MMW criteria.  For example, choosing $\eta = 0.25$ results in weights that are not monotone.  Choosing $\eta = 0.1$ results in monotone weights but use a smaller $\eta$ than the MMW weights. Consequently, this results in a few large weights and most weights being $0$ (i.e. non-robust weights). The general expression for MMW weights is provided in Theorem 3.1 below.

\begin{theorem}\label{thm.1}
For a fixed $\lambda$ and $p_m=\tau$ for all $m$, the maximum $\eta$ such that $w_{m} - w_{m^\prime}\geq 0$ for $S_{m} - S_{m^\prime}\leq 0$ for all $m,m^\prime \in \{1,\ldots,M\}$ is given by $\tilde{\eta} = S_{(1)}\sqrt{2\log\{c(\lambda)\}}$ where $S_{(1)} = \min(S_m)$, $c(\lambda) =  \lambda(1-\tau)(1-\alpha)/\{\tau(1+\lambda \alpha)\}$, and $\log\{c(\lambda)\}>0$. In this case, the thresholds are given by
\begin{equation} \label{eq.t_eta}
 \tilde t_{m}(\lambda) = \bar\Phi\left\{0.5\left(\frac{S_{(1)}\sqrt{2\log\{c(\lambda)\}}}{S_{m}}\right) + \log(c_m)\left(\frac{S_{(1)}\sqrt{2\log\{c(\lambda)\}}}{S_{m}}\right)^{-1}\right\}.
\end{equation}
\end{theorem}

\noindent A proof of this theorem is given in Section A of the Supplemental Material. 

Under the MMW criteria, the impact of weighting is minimized in that proportion of up-weighted tests is maximized among all $\eta$ values with non-increasing weights for $S_m \geq S_{(1)}$ and $p_m = \tau$. Calculating the weights for the MMW criteria is as straightforward as the fixed $\eta$ case. The MMW $\eta$ is $\tilde{\eta}= S_{(1)}(2\log\{c(\hat \lambda)\})^{1/2}$ where $\hat \lambda$ is such that  (\ref{eq.lambd}) holds when using (\ref{eq.t_eta}) for the thresholds. The value of $\tau$ guarantees that the weights satisfy the MMW criteria for non-null probabilities equal to $\tau$. This is a tuning parameter set by the investigator. As we demonstrate in our simulation studies, setting $\tau = \max(p_m)$ is effective for studies that are low in power. For studies with more robust power, we find setting $\tau$ to mean or $q$th percentile of the observed $p_m$'s can give better results.  It is important to note that weighted FDR methods control the FDR regardless of whether optimal weights are used and -- as long as weights are reasonable -- are more powerful than their un-weighted counterparts \citep{Roeder2009, Habiger2017}. 


\subsection{WABH algorithm}\label{sec.eff.size}

R code to run our proposed testing procedure along with data to replicate the results of our analysis in Section \ref{sec.data.anal} are available on GitHub \citep{McLZhe22}. A general form for implementing the WABH procedure at level $\alpha$ is given in Algorithm \ref{wabh.algorithm1}. Our implementation uses logistic regression to obtain $S_m$ and $P_m$ in step 1. The functional form between $\mbf{z}_m$ and $p_m$, and $\mbf{z}_m$ and $f_m$ is required for step 2. While our procedure only uses the estimates of $p_m$ in step 2, specifying a plausible model for $\mbf{z}_m$ and $f_m$ is important since a poor model can have a negative impact on the estimates of $p_m$.  

There are multiple measures of the impact of weighting such as the proportion of up-weighted tests (i.e., those with $w_m \geq 1$), the maximum weight, and the proportion of tests that are \textit{inconclusive} (e.g., those with $w_m < 0.1$). Inconclusive tests are those that are essentially ignored by the testing procedure, i.e., they are likely to be not rejected due to their low weight. Reporting which tests are inconclusive is an important step in implementing the WABH (or other weighting procedures) so that the tests that were essentially not included in the testing procedure can be known.

\begin{algorithm}[t]
\caption{A general form for the WABH algorithm. } \label{wabh.algorithm1}
\begin{enumerate}

    \item For $m=1,2,\ldots,M$ 
    \begin{enumerate}
        \item Compute $\bar X_m$ and $R^2_m$ to obtain $S_m$.
        \item Compute the unweighted p-value $P_m$.
    \end{enumerate}
    \item Implement AdaPT or CAMT on the unweighted p-values, extract estimates of $p_m$ for all $m$.
    \item Specify $\eta_m=\eta$ directly or specify $\tau$ and compute $\tilde\eta = S_{(1)}\sqrt{2\log\{c(\lambda)\}}$. Compute $g_m$.
    \item Compute the optimal weights by 
    \begin{enumerate}
    \item plugging $(g_m,p_m)$ into \eqref{eq.lambd} and solving for $\hat{\lambda}$, and
    \item compute optimal weight $w_m = t_m(\hat \lambda)\{M^{-1}\sum_m t_m(\hat \lambda)\}^{-1}$ using $t_m(\hat \lambda)$ in \eqref{eq.t}.
    \end{enumerate}
    \item Compute weighted p-values $Q_m = P_m/w_m$.
    \item Implement adaptive BH procedure:
    \begin{enumerate}
    \item find $k^\ast = \max\left\{m: Q_{(m)}\leq m \alpha/(\hat\pi_0 M)\right\}$ where $\hat\pi_0 = M^{-1}\sum_m (1 - p_m)$, and 
    \item compute $\delta_m(\Q) = I(Q_m\leq t_{WABH})$ where $t_{WABH} = \min\left\{\alpha, k^\ast \alpha/(\hat\pi_0 M)\right\}$.
    \end{enumerate}
    
\end{enumerate}

\end{algorithm}

\section{Simulation Study}\label{sec.sim}

To test the properties of the proposed methods we performed simulation studies with logistic regression models. We considered a two-dimensional $100 \times 100$ grid of data.  Let $\mathcal{S}_1$ index the set of false nulls. The coordinates of the tests in $\mathcal{S}_1$ were simulated from a zero-mean Gaussian random field (GRF) with $\Sigma_{s}(m,m^\prime) = \exp\{-(||\mbf{z}_m^p - \mbf{z}_{m'}^p||_2/s)^2\}$ where $\mbf{z}_m^p$ is the two-dimensional coordinates for point $m$ and $||\cdot||_2$ is the $l_2$-norm.  The tests in $\mathcal{S}_1$ were those with the largest $K = ||\mathcal{S}_1||$ simulated values. The data were generated via
\begin{equation}\label{eq.sim.study}
\mbox{logit}\{\Pr(X_{im}=1|Y_i,b_i)\} = \alpha^\ast_0 + \alpha^\ast_{0m} + \alpha^\ast_{1m}Y_i+ b_i,
\end{equation}
where $\alpha^\ast_0 = -1$, $\alpha^\ast_{0m}$ follows a zero-mean GRF with covariance function $C^{2} \Sigma_{50}(m,m^\prime)$, $\alpha^\ast_{1m}\sim U(0,2\theta)$ for all $m \in \mathcal{S}_1$ ($\alpha^\ast_{1m}=0$ otherwise), $b_i \sim N(0,0.8^2)$, and $Y_i = 0.5b_i + \epsilon_i$ where $\epsilon_i \sim N(0,0.5^2)$. Recall from (\ref{eq.logreg.SE}), that the power heterogeneity is driven by $Var(X_{im})=\bar X_m(1-\bar X_m)$, which will be a function of $\alpha^\ast_{0m}$ in that extreme $\alpha^\ast_{0m}$ will have low $Var(X_{im})$.  Thus, $C^{2}=Var(\alpha^\ast_{0m})$ controls the amount of power heterogeneity among the tests. 

The simulation settings were varied over $C=0.5$, $1.5$, and $3$, corresponding to low, moderate, and high power heterogeneity, respectively, $s=0.01$, $5$, and $10$, corresponding to low, moderate, and high spatial clustering, respectively, $K=100$ or $500$, and the expected effect size $\theta = 0.25,0.5,$ or $0.75$ for low, moderate, or high average power. All simulations used $n=200$. The fitted model for the $m$th test was
$$\mbox{logit}\{\Pr(X_{im}=1|Y_i,X_{i}^+)\} = \alpha_{0m} + \alpha_{1m}Y_i + \alpha_{2m}X_{im}^+  $$
where $X_{im}^+ = \mbox{logit}\{(M-1)^{-1}\sum_{j \neq m} X_{ij}\}$. Section B of the Supplemental Material contains example plots of the data.

\begin{figure}[t]
\begin{center}
\includegraphics[width= 6.4in, height=4.2in]{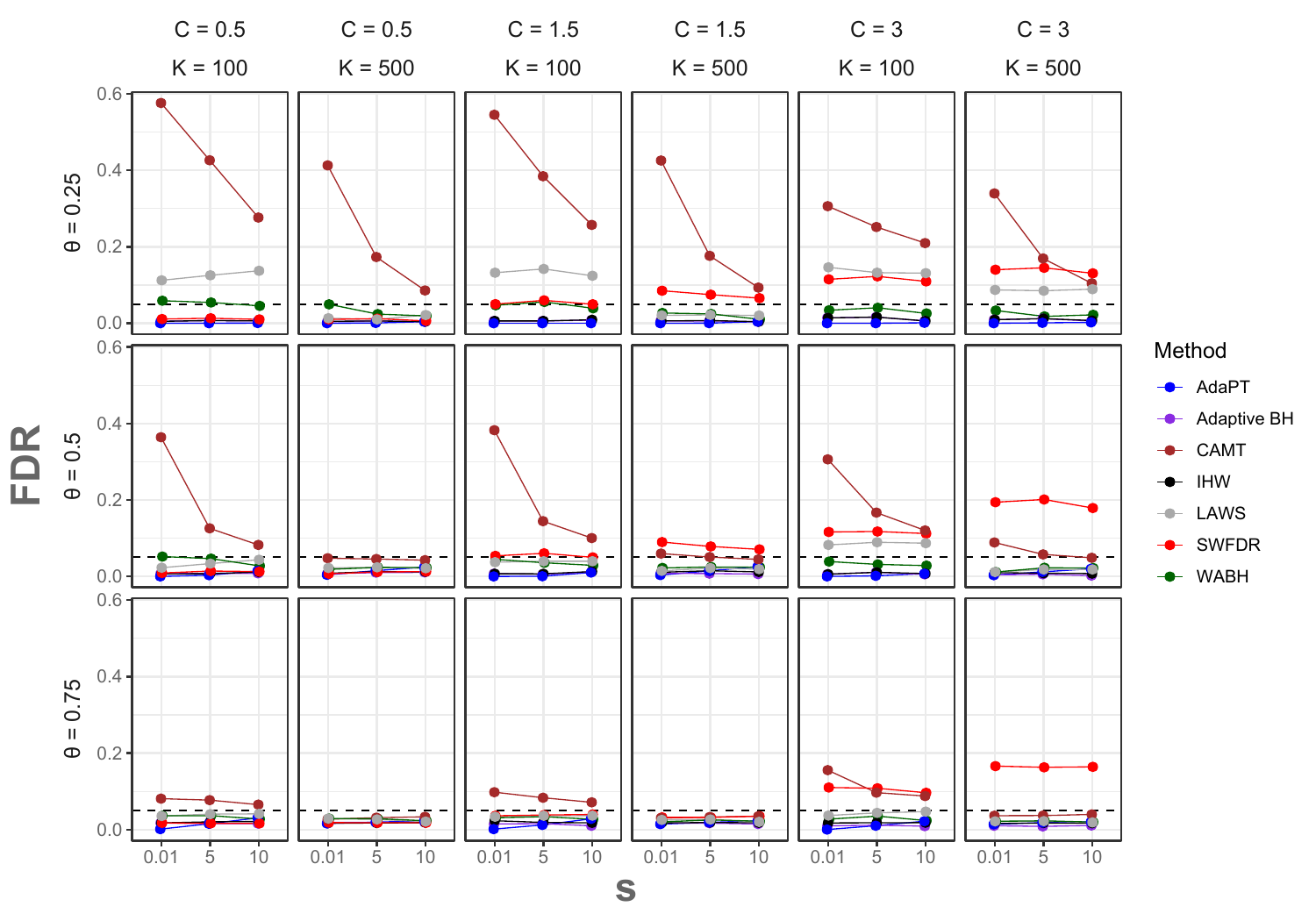} 
\end{center}
\vspace{-.2in}
\caption{Average false discovery rates (FDR) by power heterogeneity ($C$), number of true signals ($K$), effect size ($\theta$), and spatial dependence ($s$) for $M=10000$ tests and $n=200$.}\label{fig.bin.sim1}
\end{figure}

For all settings, the p-value weights were estimated using Algorithm \ref{wabh.algorithm1} where $p_m$ was estimated using AdaPT, or CAMT. For both AdaPT and CAMT, 
\begin{equation}\label{eq.pm.model}
\log\{p_m/(1-p_m)\} = \gamma_0  + \gamma_1 h_5(z_{m1}^p) + \gamma_2 h_5(z_{m2}^p) + \gamma_3 h_5(z_{m1}^p z_{m2}^p) 
\end{equation}
where $z_{mj}^p$ is coordinate $j$ of test $m$ and $h_5(z)$ is a vector of $5$ degree natural cubic splines with evenly spaced knots evaluated at $z$. Further, $\log(k_m) = h_5(S_m)$. To select $\eta$ we used the MMW criteria with $\tau = 0.5$ or $0.9$. We note that for AdaPT and CAMT the the beta assumption on the distribution of non-null p-values is likely misspecified. We also included the Adaptive BH, IHW, LAWS, and SWFDR procedures in the simulation. Adaptive BH was implemented with $\hat\pi_0$ being estimated using \cite{Sto07} with a threshold set at $0.05$ as suggested for dependent data \citep{Blanchard2009}. IHW was fitted with one covariate ($S_m$), as this was all the software allowed, with five-folds and automatic selection of the number of bins. SWFDR was fitted with a design matrix consisting of the two-dimensional coordinates and $S_m$ to estimate the null probability for each test. To fit LAWS we used a threshold of $0.9$ (the default) with a Gaussian kernel and bandwidth set to 4.5. LAWS does not model the non-null p-value distribution via covariates, thus $S_m$ was not used.

The methods are compared in terms of $FDR = B^{-1}\sum_b I(R_b>0)(V_b/R_b)$ and $Power = ETP/K$ ($ETP = B^{-1} \sum_b S_b$) where $S_b$, $V_b$ and $R_b$ denote the number of correct discoveries, false discoveries and total discoveries from the $b$th iteration. Procedures set $FDR$ control level to $\alpha=0.05$. All simulations were run for $B=500$ iterations.  For brevity, we show the WABH when $p_m$ was estimated using CAMT with $\tau=0.9$ only. Section B of the Supplemental Material contains results of WABH with AdaPT and other $\tau$ values along with other common methods.

\begin{figure}[t]
\begin{center}
\includegraphics[width= 6.4in, height=4.2in]{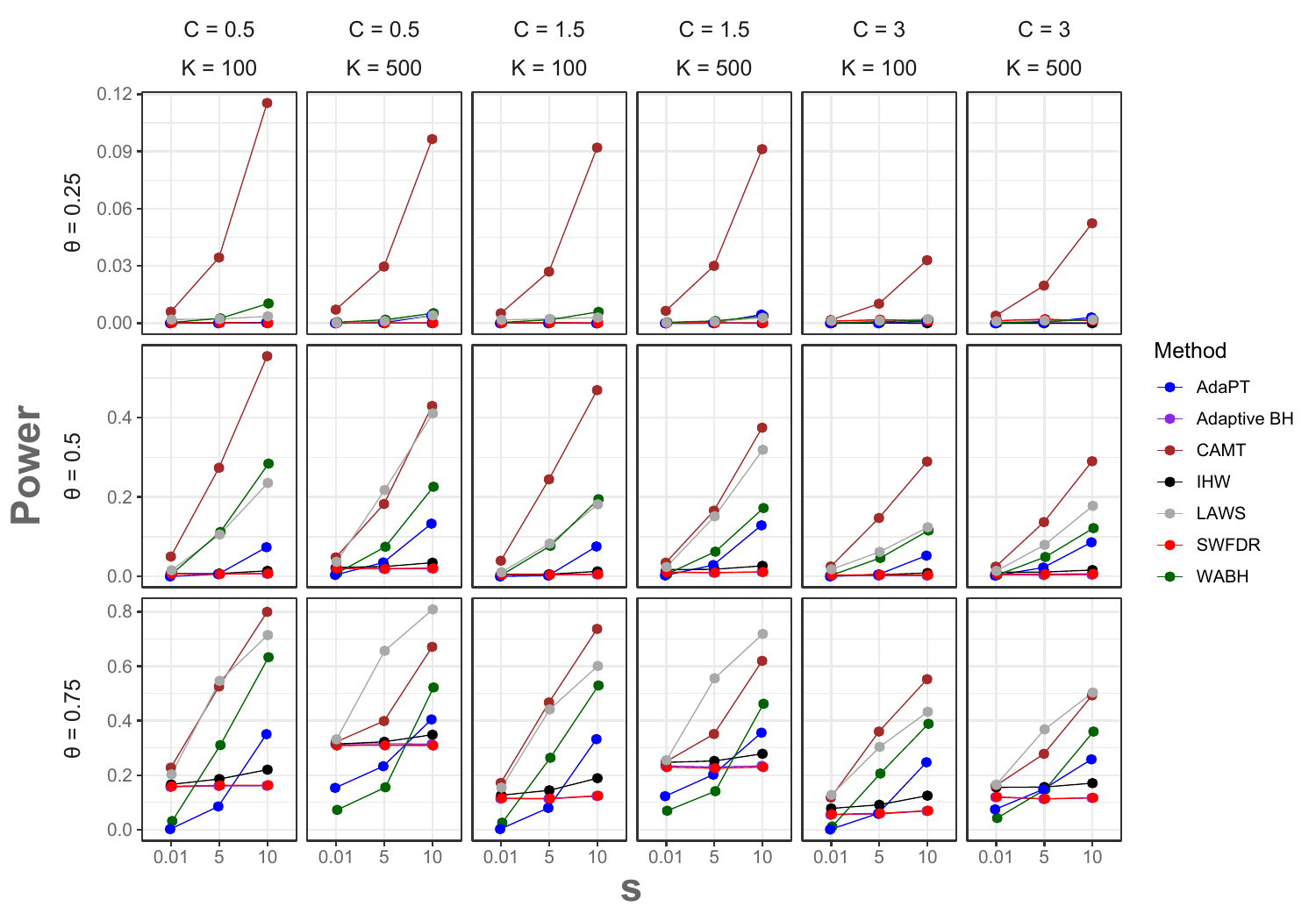} 
\end{center}
\vspace{-.2in}
\caption{Estimated power ($ETP/K$) by the power heterogeneity ($C$), number of true signals ($K$), effect size ($\theta$), and spatial dependence ($s$) for $M=10000$ tests and $n=200$.}\label{fig.bin.sim2}
\end{figure}

In Figure \ref{fig.bin.sim1}, we present summarized FDR results of simulation studies. The AdaPT, Adaptive BH, IHW, and WABH procedures all controlled the FDR with values that are less than or close to the nominal level in all settings. The CAMT had high $FDR$ when the proportion of false null hypotheses or expected effect size ($\theta$) was low. For example, when $K=100$ and $\theta=0.25$, the $FDR$ ranged from $0.2-0.6$ and $0.2-0.3$ in the low to moderate, and high heterogeneity settings, respectively. For $K=100$ and $\theta=0.75$, the CAMT $FDR$ ranged from $0.06-0.16$, while for $K=500$ the FDR was controlled reasonably well for $\theta \geq 0.5$. The LAWS procedure also resulted in $FDR$ values that were above the nominal level when $\theta=0.25$, and some $\theta=0.5$ settings. Lastly, the $FDR$ for the SWFDR procedure was above the nominal level in all high heterogeneity settings.

In Figure \ref{fig.bin.sim2}, we present summarized power results of simulation studies. Overall, among the procedures which control $FDR$ in most settings (i.e., AdaPT, Adaptive BH, IHW, and WABH) the WABH procedure had the largest power. AdaPT had larger power for $\theta = 0.75$, $K=500$ when $s \leq 5$. For $\theta=0.75$, the LAWS procedure controlled the FDR and had the largest power (CAMT has larger power only when the FDR$\geq \alpha$). Further, LAWS performed well (FDR controlled and high power) for $\theta=0.5$ and $K=500$. As a result, the LAWS procedure appears to perform well in high-powered settings, however, such settings are not likely in our application of interest.

\section{Data Analysis}\label{sec.data.anal}

Our sample was drawn from the population described in detail by \cite{Youetal15}, and follows the same inclusion/exclusion criteria, behavioral testing, and behavioral analyses. In brief, all participants were adults with chronic left-hemisphere stroke and aphasia. All individuals were scanned using a 3T MRI scanner. Lesions were obtained by hand by an expert on a high-resolution T2-scan for optimal identification of lesion boundaries. Lesions were coregistered to the individual's T1 scan. Each individual’s lesion was warped to have a common size and shape through enantiomorphic normalization \citep{Nacetal08} using our clinical toolbox \citep{Roretal12}. Therefore, for each individual, the lesion was mapped as a binomial volume with a resolution of $181\times 217 \times 181$ voxels (each 1mm$^3$), though many of these voxels are outside the brain or have zero damage for all subjects. For this study, we included the $220$ individuals enrolled at the time of these analyses. Data from $834582$ \textit{candidate voxels} -- voxels with damage to at least one subject -- were considered. 

To fit WABH we used CAMT to estimate $p_m$ with the MMW criteria and $\tau = 0.9$. The side-information consists of $\mbf{z}_m^p$ the $3$-dimensional voxel coordinates and $S_m$. The relationship between $\mbf{z}_m^p$ and $p_m$ was similar to \eqref{eq.pm.model} with natural cubic splines on all three coordinates and all two-way interactions. We used 12 degrees of freedom for the splines which had the smallest BIC among the many values tested. The comparison methods included the 10\% Rule, BH, ABH, AdaPT, and CAMT methods. For AdaPT, penalized regression splines with an automatic degree of freedom selection were used to relate $\mbf{z}_m^p$ and $p_m$ and $S_m$ with $f_m$. We attempted an analysis with LAWS on this data also but it was not computationally feasible and failed to converge. Part of the issue is that LAWS requires a full cubic 3D structure. After removing slices with no candidate voxels more than $2\times 10^6$ tests were still present (more than twice the tests of other procedures). For AdaPT and the WABH with AdaPT, we removed voxels with damage in less than 0.5\% of subjects due to the convergence issue. Among voxels with non-zero damage, $360038$ ($43.1\%$) have damage in less than 5\% of subjects. Codes and data to replicate the data analyses are available on GitHub \cite{McLZhe22}.

\begin{figure}[t]
\centering
\begin{tabular}{ccccccc}
\includegraphics[scale=0.16]{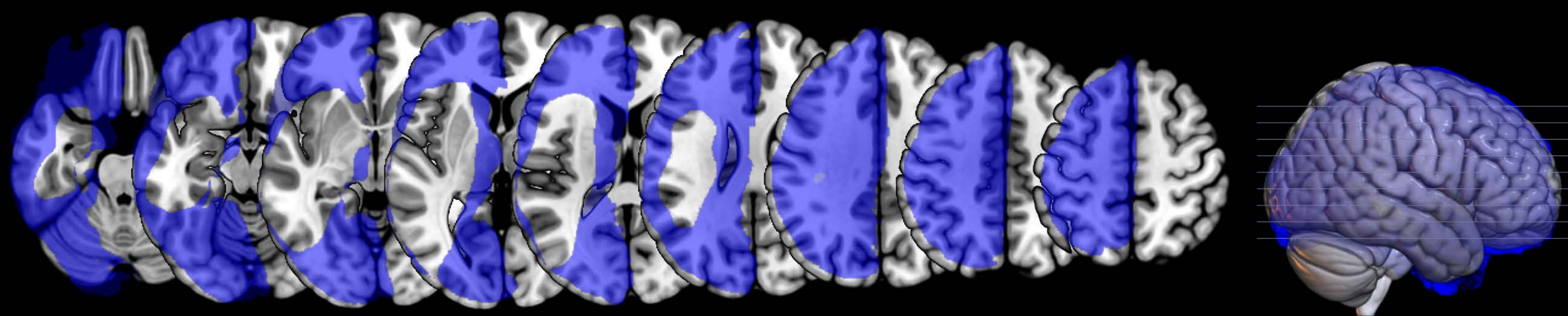} \\
\includegraphics[scale=0.16]{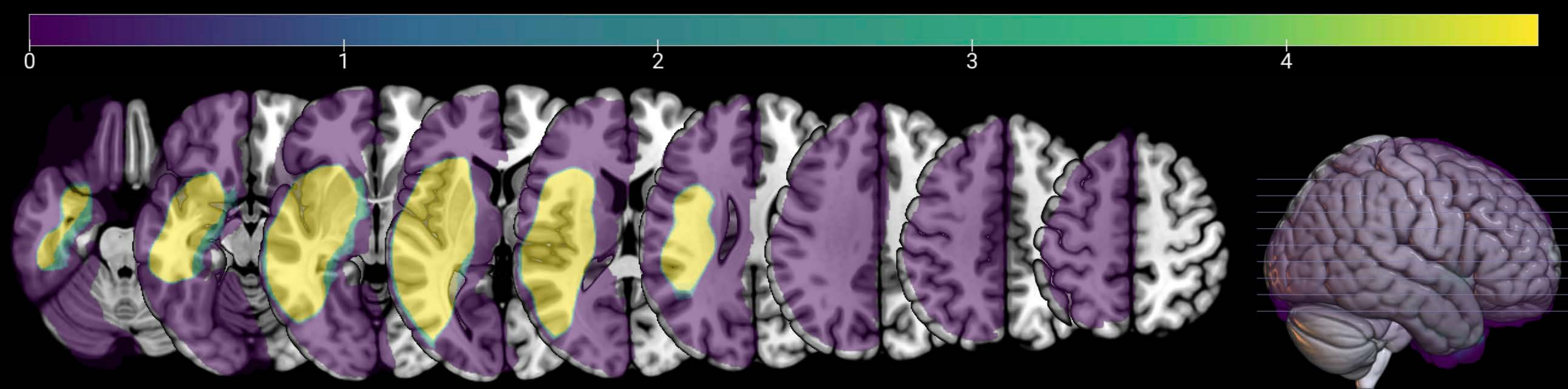} \\
\includegraphics[scale=0.16]{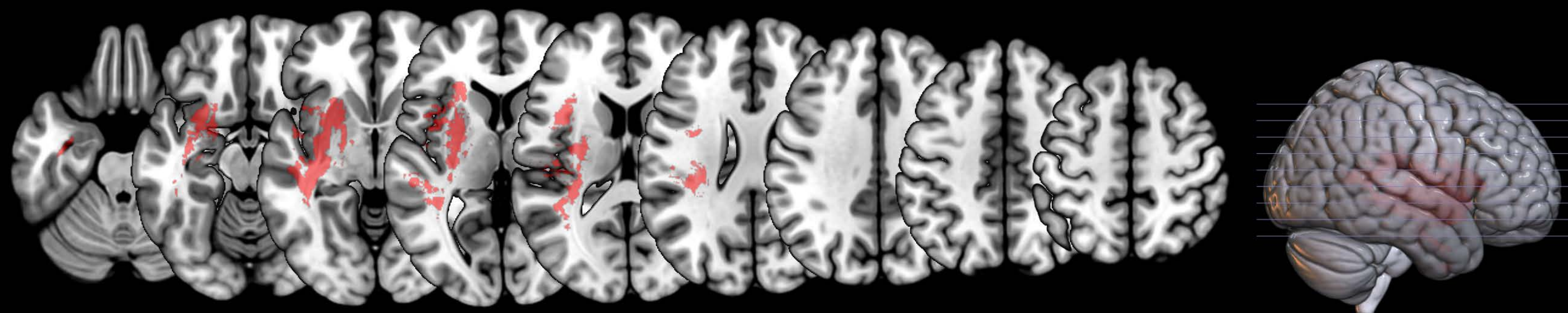} \\
\end{tabular}
\caption{Inconclusive voxels (top, in blue), p-value weights (middle), and significant voxels (bottom, in red) for the WABH procedure.  The plots are overlayed on a white structural brain image for reference.}\label{fig.data.weights}
\end{figure}

Lesion status was regressed as a function of the Aphasia Quotient (AQ) score ($Y$) and total lesion size ($X^+$).  Checking the assumptions of all tests individually is not possible, so scatterplots of $Y$ and $X^+$ were examined to determine the nature of the relationship between $Y$ and $X_{m}$.  This should suffice since $X^+$ is a measure of the average probability, and if the average probability is related to some transformation of $Y$ then it is plausible that the voxel-level probabilities are related to the same transformation. Scatterplots of $Y$ and $X^+$ showed a linear relationship when both were logit transformed. As a result, both $Y$ and $X^+$ were included in the model after a logit transformation.

In Figure \ref{fig.data.weights}, we present a map of the estimated p-value weights, inconclusive voxels, and significant voxels for the WABH procedure with prior non-null probability estimated using CAMT.  By knowing the weights, voxels, where the tests were inconclusive, can be shown. For example, the blue voxels in Figure \ref{fig.data.weights} have weights less than $0.1$ and thus could very likely contain type II errors. Such results are critical to show researchers which areas still require further study. Other weight metrics show that the WABH with AdaPT resulted in 36\% up-weighted tests, 64\% inconclusive tests, and $2.79$ as the maximum weight. The WABH procedure with CAMT has 23\% up-weighted tests, 75\% inconclusive tests, and $4.84$ as the maximum weight. 

The WABH with CAMT, AdaPT, and CAMT find $26568$, $20$, and $174953$ significant voxels respectively, while WABH with AdaPT, 10\% Rule, BH, and ABH find no significant voxels. Many of the significant voxels for WABH with CAMT appear to be located in and around the inferior frontal gyrus, which contains Broca's region which is a main area linked to speech production.

\section{Discussion}\label{sec.disc}

In this paper, we proposed the use of weighted adaptive BH hypothesis testing for VLSM analysis. While the weighted adaptive BH procedure has been proposed by others, this paper was the first to incorporate heterogeneous prior non-null probability and proposed approach for estimating effect sizes in a manner consistent with anticipated low power assumptions of VLSM (see Theorem 3.1). The specific weighting scheme is available in Algorithm 1. Our simulation studies showed that our proposed method has a better performance than the other commonly used methods. Specifically, we found that while CAMT has high power the $FDR$ was well above the nominal level, particularly in settings with a smaller expected effect size ($\theta$) or a number of non-nulls ($K$). LAWS also had difficulty controlling the FDR for low values of $\theta$. An in-depth assessment of why these methods -- both of which have solid theoretical guarantees on their FDR values -- fail to control the FDR is beyond the scope of this paper. However, it is evident that these methods have difficulty when $\theta$ and $K$ are small. As a result, their inflated FDRs are in situations where estimating properties about $p_m$ and $f_m$ are challenging due to low power and/or a small number of non-null tests. The proposed method was the most powerful among those that controlled the FDR for most settings. The findings of the data analysis are consistent (though can't be confirmed) with the findings of the simulation study, where CAMT (AdaPT) found many more (less) significant voxels than our proposed method.

Our proposed WABH procedure ignored the dependence between the hypotheses tests. The WABH procedure has been shown to provide asymptotic FDR control under a weak dependence structure on the p-values \citep{Habiger2017}, which is plausible for our setting. \cite{Benjamini2001} showed that the BH procedure still controls the FDR under a positive regression dependence structure (PRDS) and proposed modifications to the original BH procedure for arbitrary dependence. The PRDS property is satisfied if the test statistics are Gaussian, nonnegatively correlated and the testing hypotheses are one-sided. Since VLSM are usually one-sided tests and the spatial correlation between the test statistics will be (mostly) non-negative, the assumptions proposed by \cite{Benjamini2001} may be reasonable for the application of interest. However, extending the WABH to more general dependence scenarios is of interest.

In the data analysis, the number of discoveries was positively associated with the severity of weighting (i.e., heavier weighting resulted in more discoveries). However, heavy weighting results in many inconclusive hypotheses (up to 75\% in our data analysis), and the regions are likely to include type II errors which need to be studied further. P-value weighting results in more discovered voxels by down-weighting voxels where discovery isn't likely and up-weighting voxels where it is. The result is more overall power in exchange for essentially not testing some voxels. It is important to acknowledge these later regions in reporting. This is why results such as Figure \ref{fig.data.weights} should be included when they are employed so that the impact of weighting is transparent. Codes and data to replicate the data analyses are available on GitHub \cite{McLZhe22}.

\section*{Acknowledgements}
The authors gratefully acknowledge that this research has been supported by the National Institutes of Health grant (R01 DC009571) to CR and JF.

\bibliographystyle{asa}
\bibliography{WABH}

\newpage

\begin{center}
\textbf{\LARGE Supplemental Material}
    
\end{center}

\section*{A. Proof of Theorem 3.1}

For a fixed $p_m=p$ for all $m$, to prove theorem 1 we wish to find the maximum $\eta$ such that
\begin{equation}\label{cond.1}
 w_{m} - w_{m^\prime}\geq 0\mbox{ for }S_{m} - S_{m^\prime}\leq 0\mbox{ for all }m,m^\prime \in \{1,\ldots,M\}.
\end{equation}
To show that $\tilde \eta$ in the text is the unique solution to (\ref{cond.1}), we first establish sufficient criteria for the weights to be non-increasing in $S$, then show that $\tilde \eta$ is the largest such $\eta$ that satisfy this condition.

Let $t_m(\lambda) \equiv t(\lambda,\eta,S_{m},p_m)$ where recall that 
$$ t(\lambda,\eta,S_{m},p_m) = \bar\Phi\left[0.5g_m + \log\{c(\lambda;p_m)\}g_m^{-1}\right]$$ 
where
$$ c(\lambda;p_m) = \frac{\lambda(1-p_m)(1-\alpha)}{p_m(1+\lambda\alpha)}$$
For a fixed $p_k = \tau$ for all $k$, note that $w_{m} - w_{m^\prime}\geq 0 \Leftrightarrow t(\lambda,\eta,S_{m},\tau) \geq  t(\lambda,\eta,S_{m^\prime},\tau)$.  Thus, we seek the largest $\eta$ such that $t^\prime(c,\eta,S_m,\tau)\leq 0$ for all $S_m\in[S_{(1)},S_{(M)}]$, where $S_{(1)} = \min(S_m)$, $S_{(M)} = \max(S_m)$ and
\begin{eqnarray}\label{t.S.der}
t^\prime(c,\eta,S_m,\tau) &=& \frac{d}{dS} t(c,\eta,S,\tau) \bigg|_{S=S_m} \nonumber \\
&=& -\phi \left[\left(\frac{\eta}{2S_m}\right) + \log\{c(\lambda;\tau)\} \left(\frac{S_m}{\eta}\right)  \right]\left[\frac{\log\{c(\lambda;\tau)\}}{\eta}   - \frac{\eta}{2S_m^2}\right].
\end{eqnarray}
Note that, 
\begin{enumerate}
\item[(i)] if $\eta_m=S_m \sqrt{2\log\{c(\lambda;\tau)\}}$ then $t^\prime(c,\eta_m,S_m,\tau)=0$. 
\end{enumerate}
Further, the derivative of the latter portion of (\ref{t.S.der}) with respect to $\eta$ is negative for all $\eta>0$ (which is sufficient since the first portion is negative for all $p$, $S$ and $\eta$), thus for $\eta_m=S_m \sqrt{2\log\{c(\lambda;\tau)\}}$
\begin{enumerate}
\item[(ii)] $t^\prime(\lambda,\eta_m-\epsilon,S_m,\tau)<0$ for $0<\epsilon<\eta_m$, and
\item[(iii)] $t^\prime(\lambda,\eta_m+\epsilon,S_m,\tau)>0$ for $\epsilon>0$.
\end{enumerate}  
By (i) if $S_{m} - S_{m^\prime} < 0$ then $\eta_m - \eta_{m \prime} < 0$ and by (ii) $t^\prime(c,\eta_{m \prime},S_m,\tau)<0$.  As a result, $t^\prime(c,\tilde \eta,S_m,\tau)\leq 0$ for all $S_m\in[S_{(1)},S_{(M)}]$ where $\tilde \eta =S_{(1)} \sqrt{2\log\{c(\lambda;\tau)\}}$.  As a result, (\ref{cond.1}) holds for $\tilde \eta$.  The fact that $\tilde \eta$ is the largest $\eta$ to satisfy (\ref{cond.1}) follows from (iii).

\section*{B. Simulation Study Results}

\begin{figure}[t]
\begin{center}
\includegraphics[width= 6.4in, height=4.2in]{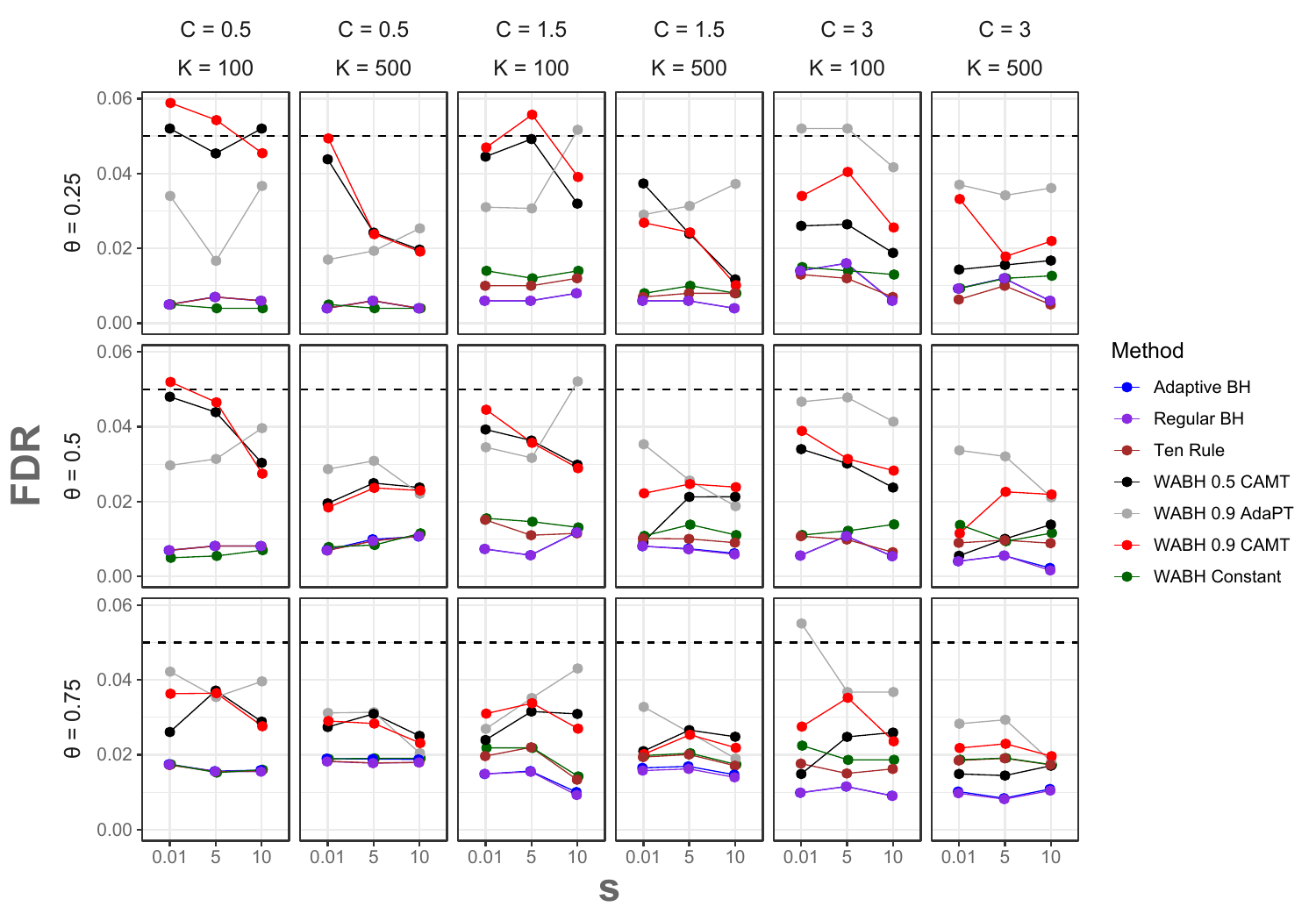} 
\end{center}
\vspace{-.4in}
\caption{Average false discovery rates (FDR) by power heterogeneity ($C$), number of true signals ($K$), effect size ($\theta$), and spatial dependence ($s$) for $M=10000$ tests and $n=200$.}\label{fig.bin.simfdr2}
\end{figure}

\begin{figure}[t]
\begin{center}
\includegraphics[width= 6.4in, height=4.2in]{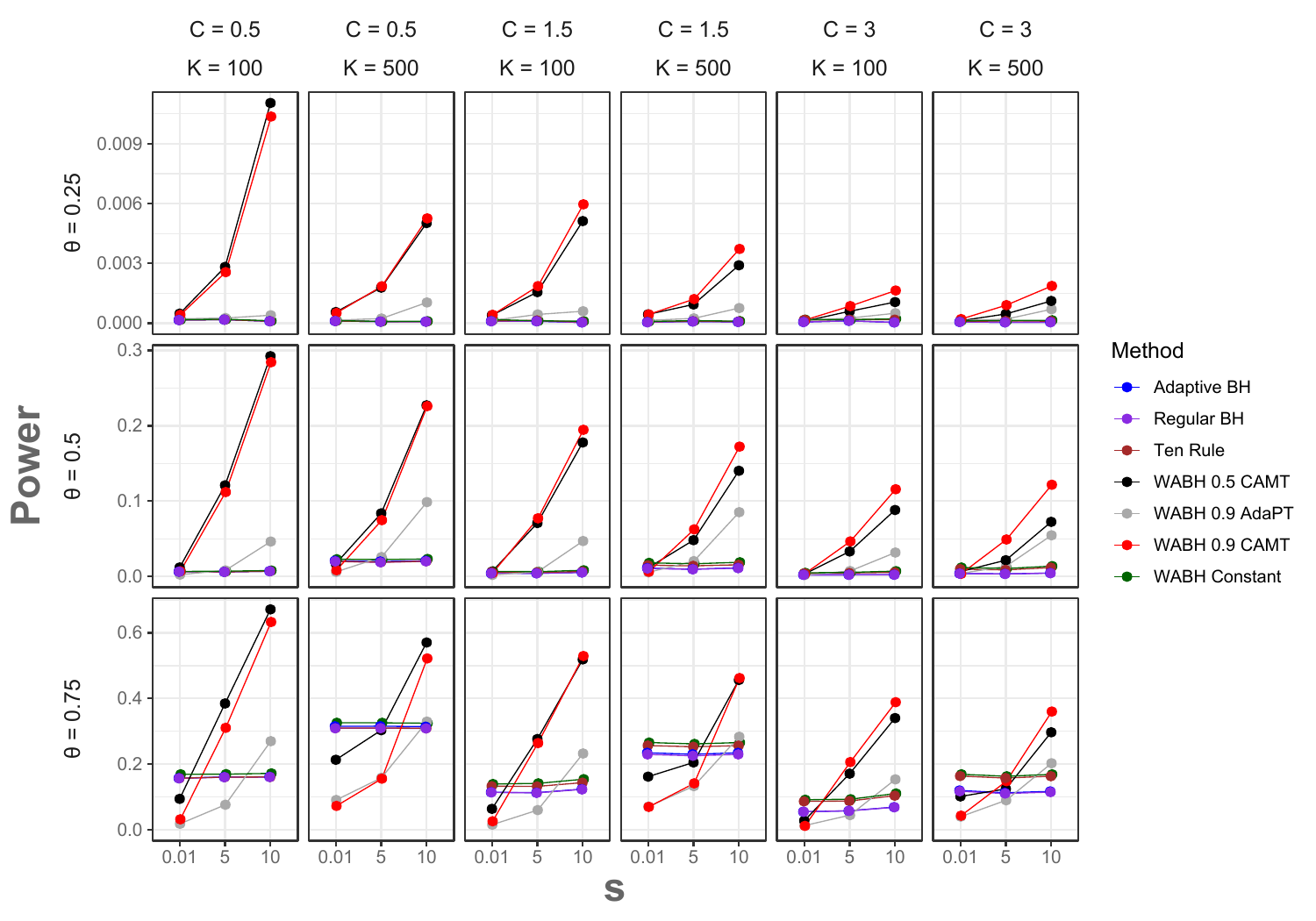} 
\end{center}
\vspace{-.4in}
\caption{Estimated power ($ETP/K$) by the power heterogeneity ($C$), number of true signals ($K$), effect size ($\theta$), and spatial dependence ($s$) for $M=10000$ tests and $n=200$.}\label{fig.bin.simpower2}
\end{figure}

\begin{figure}[t]
\centering
\begin{tabular}{ccc}
 \includegraphics[scale=0.35]{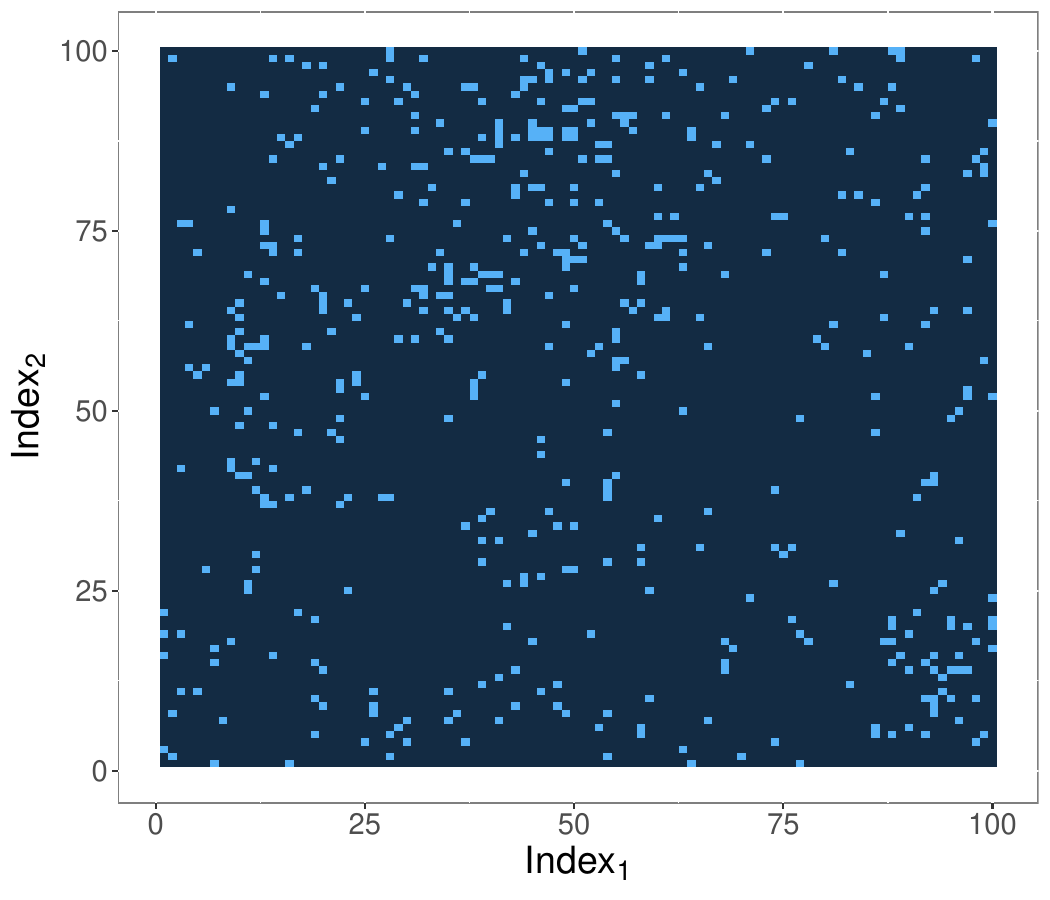} & \includegraphics[scale=0.35]{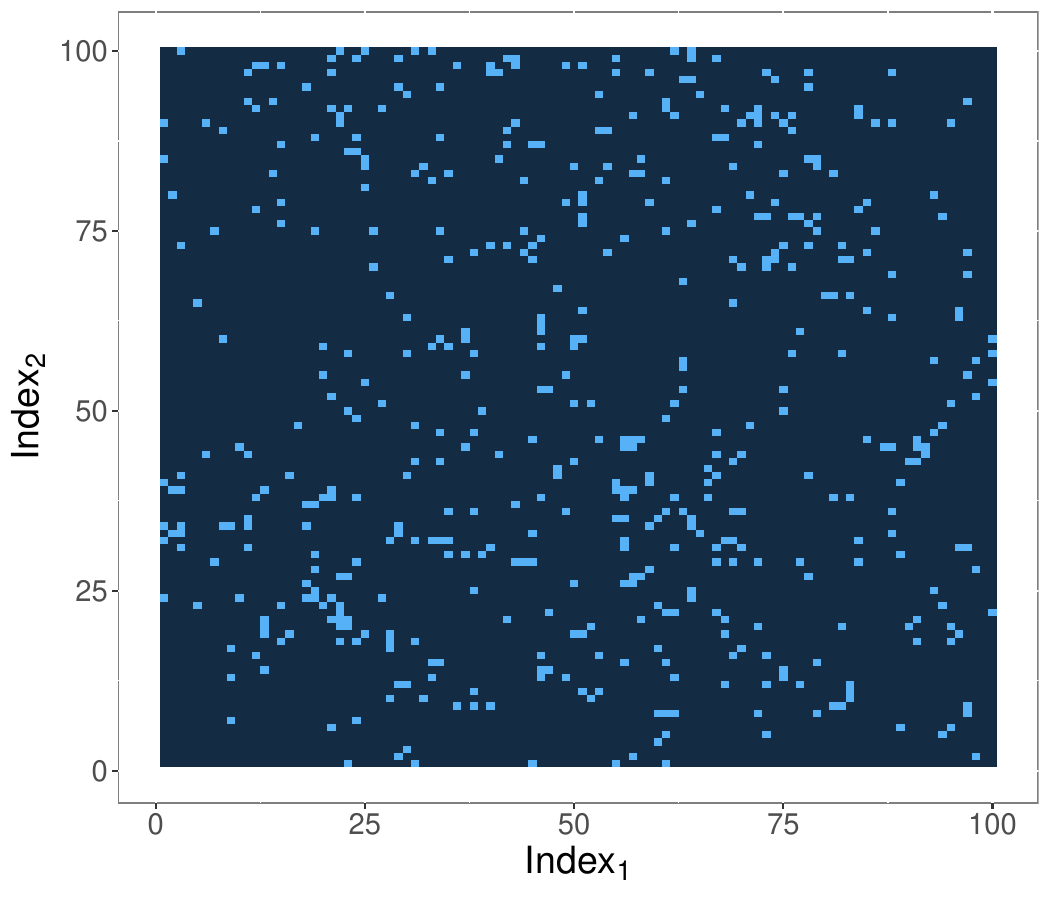}\\
 \includegraphics[scale=0.35]{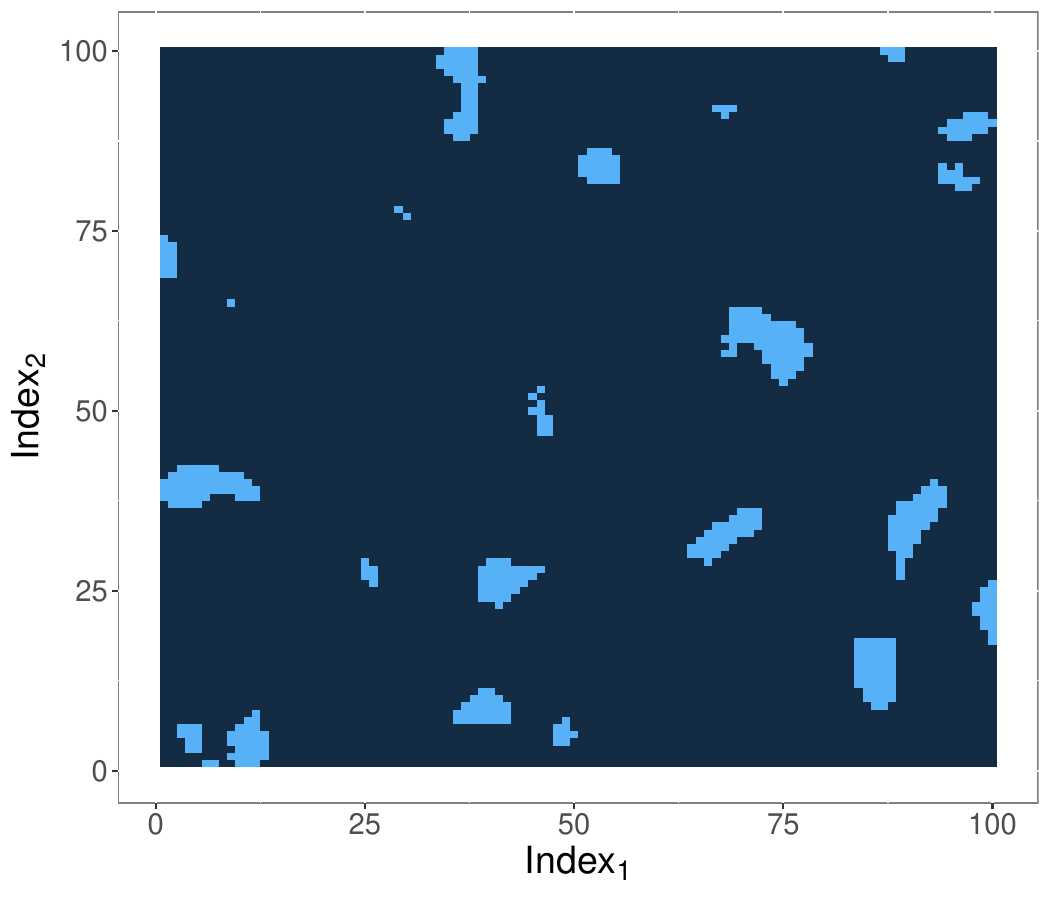} & \includegraphics[scale=0.35]{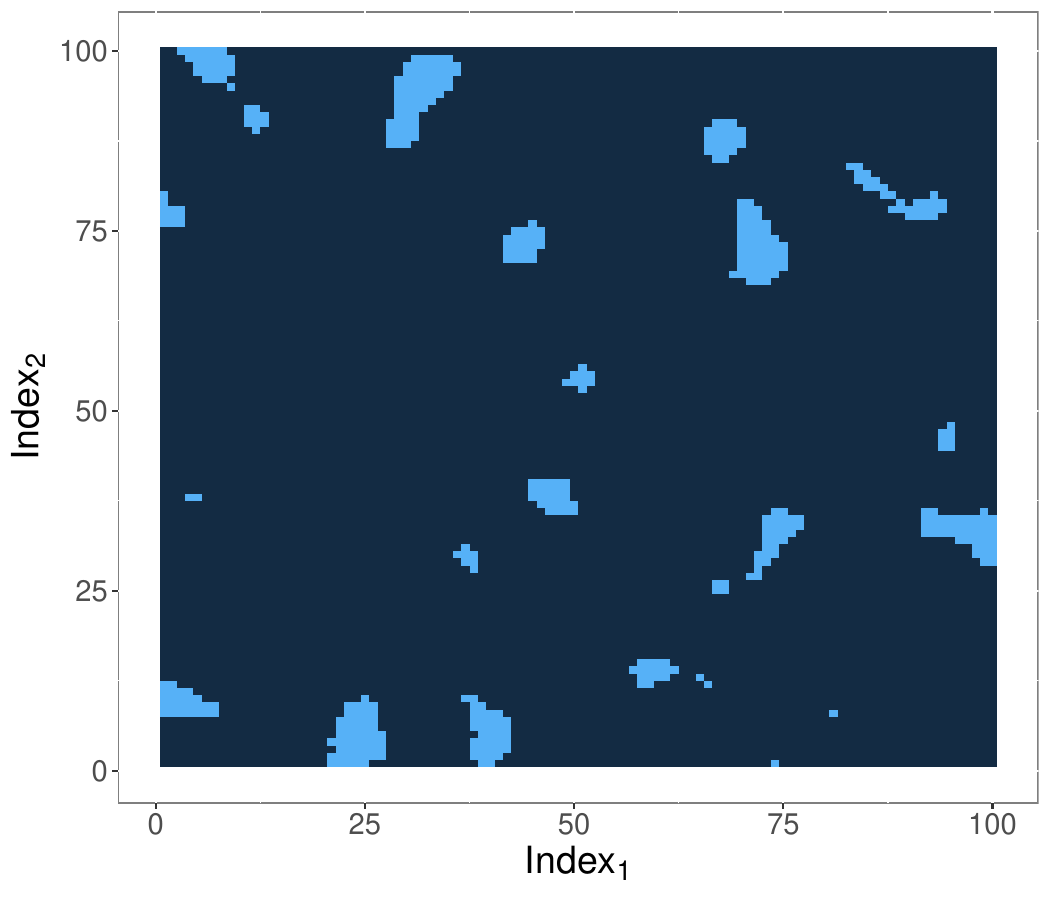}\\
\includegraphics[scale=0.35]{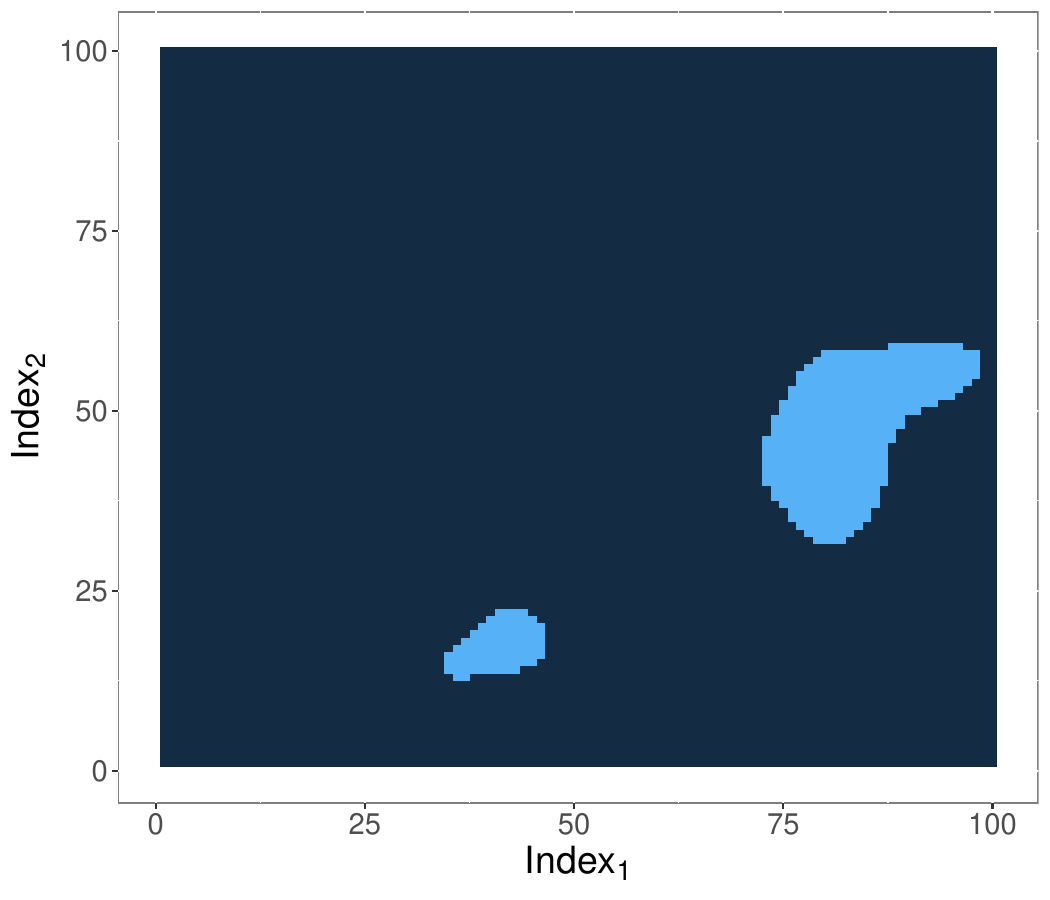} & \includegraphics[scale=0.35]{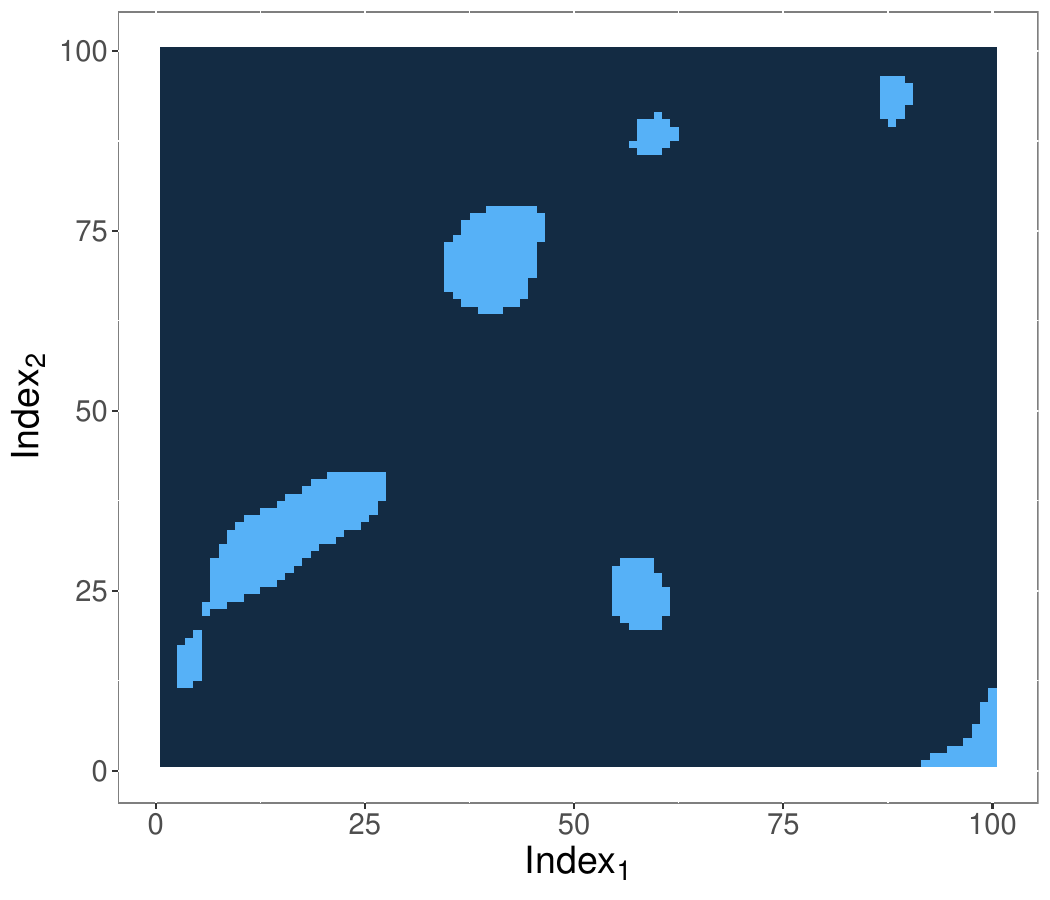}
\end{tabular}
\caption{Examples of simulation signals (light blue) with low to high spatial dependence (top to bottom) for 500 true signals out of $M=10000$ tests.}\label{fig.signal.example}
\end{figure}

\begin{figure}[t]
\centering
\begin{tabular}{ccc}
 \includegraphics[scale=0.35]{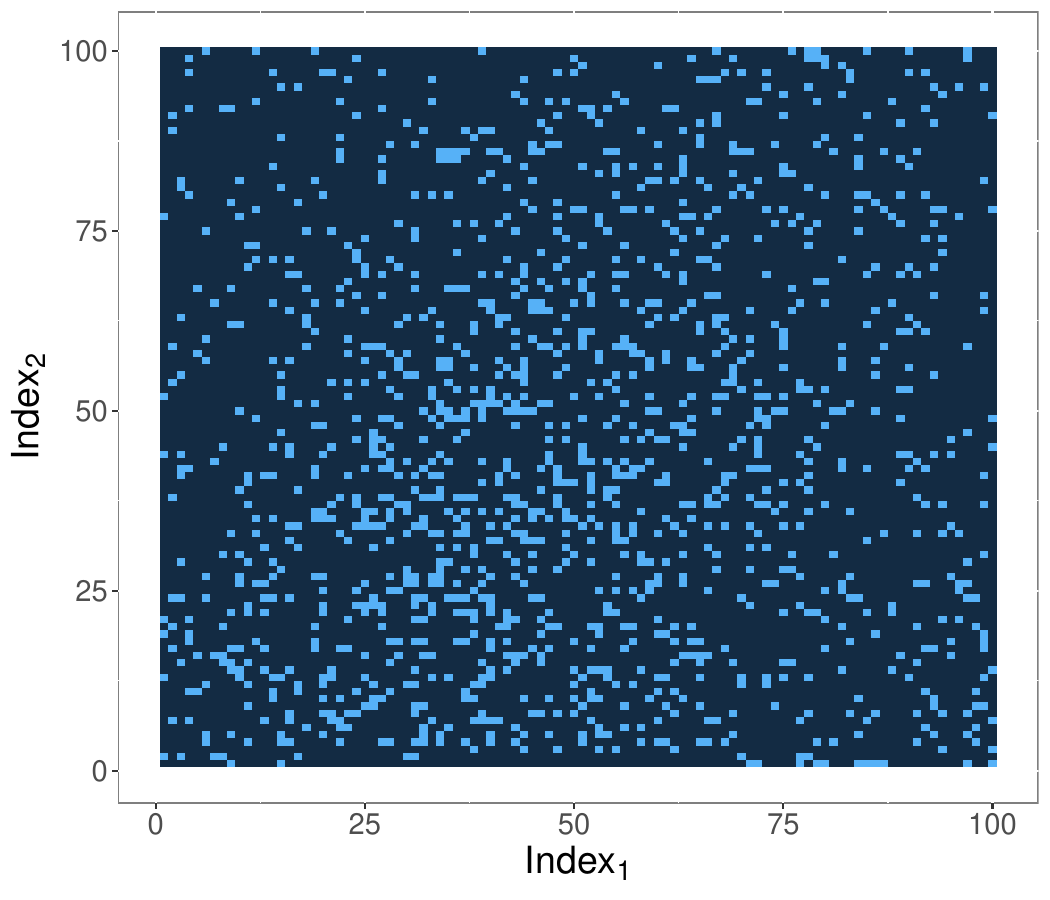} & \includegraphics[scale=0.35]{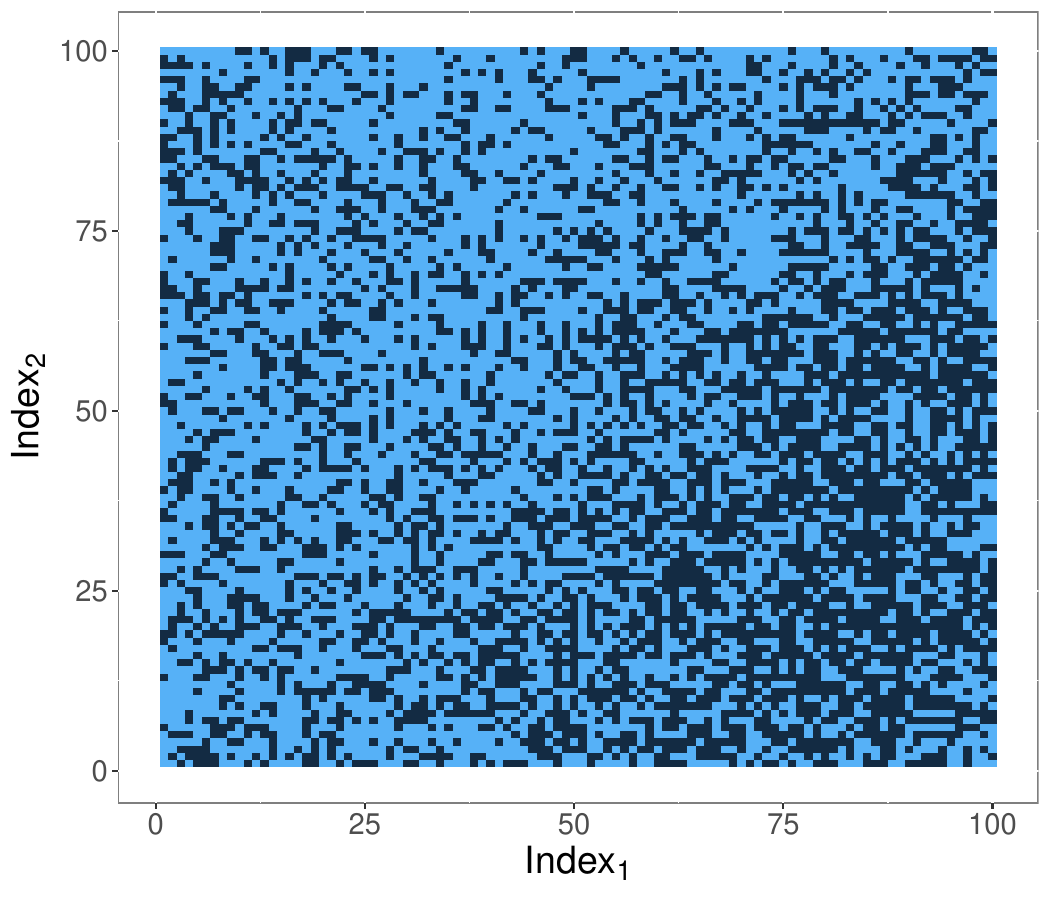}\\
 \includegraphics[scale=0.35]{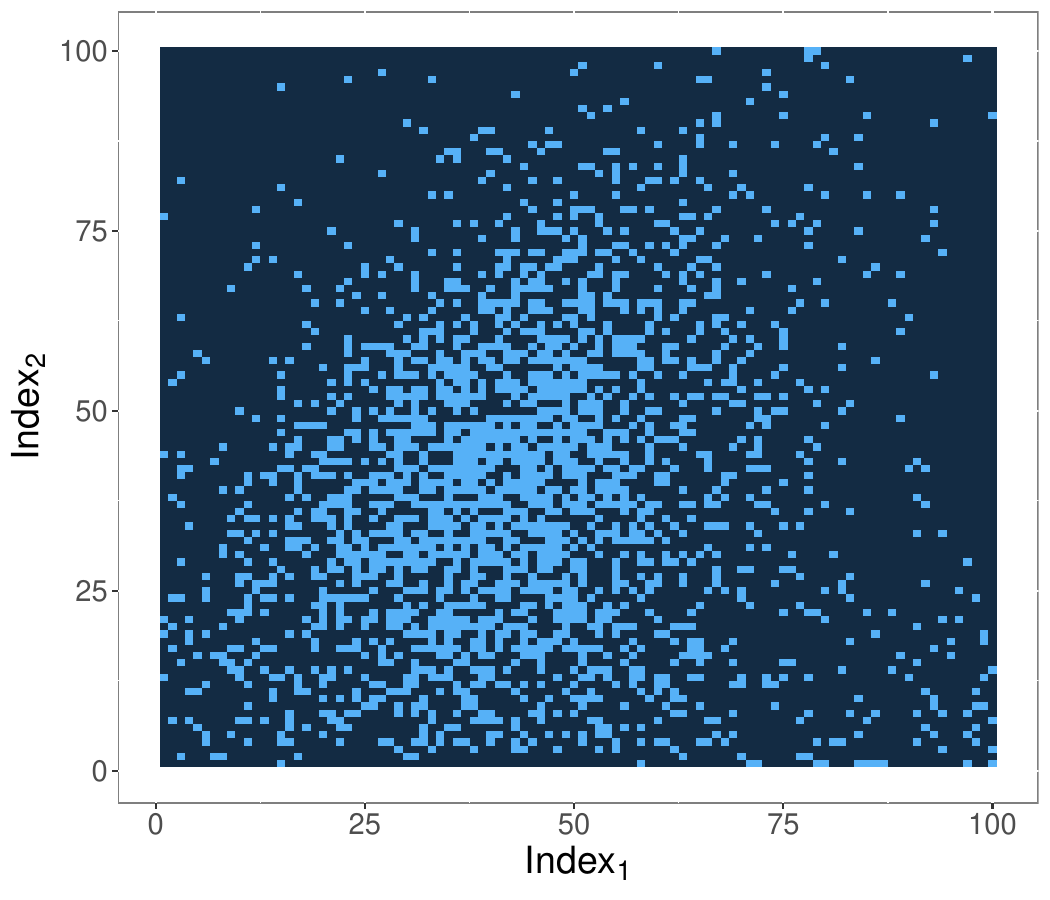} & \includegraphics[scale=0.35]{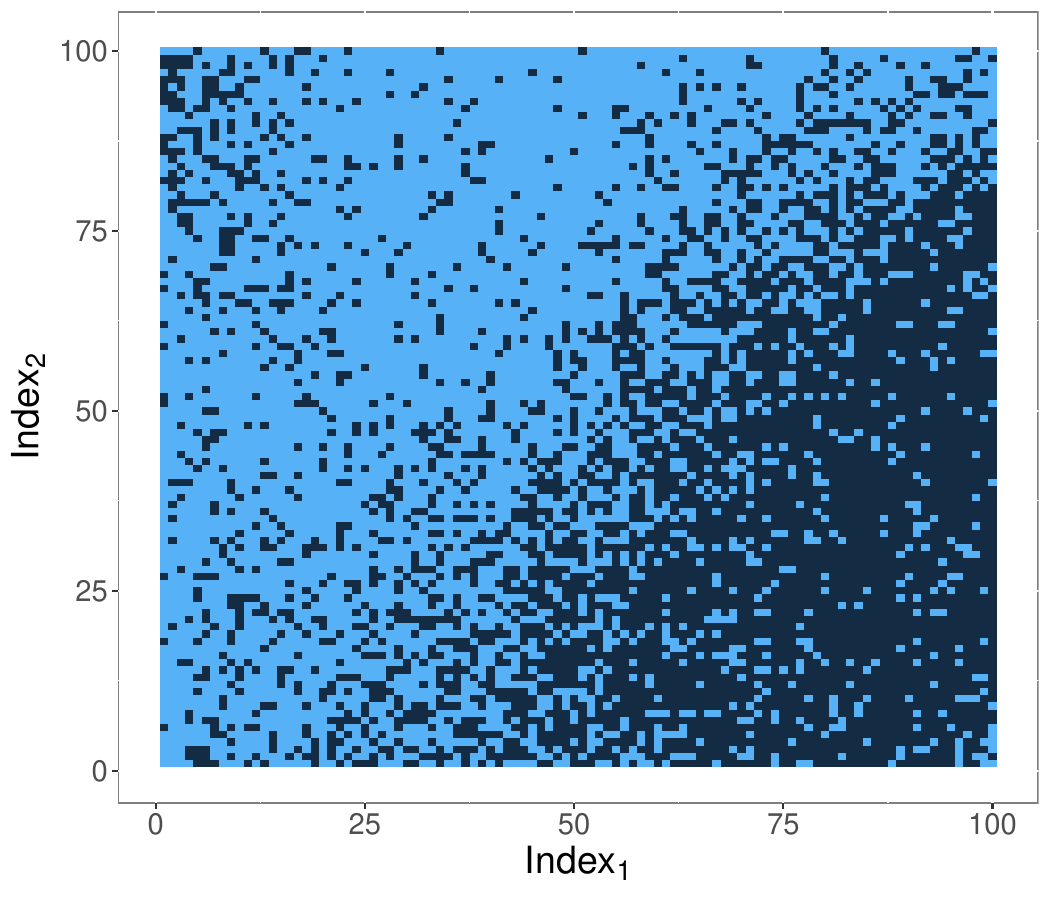}\\
\includegraphics[scale=0.35]{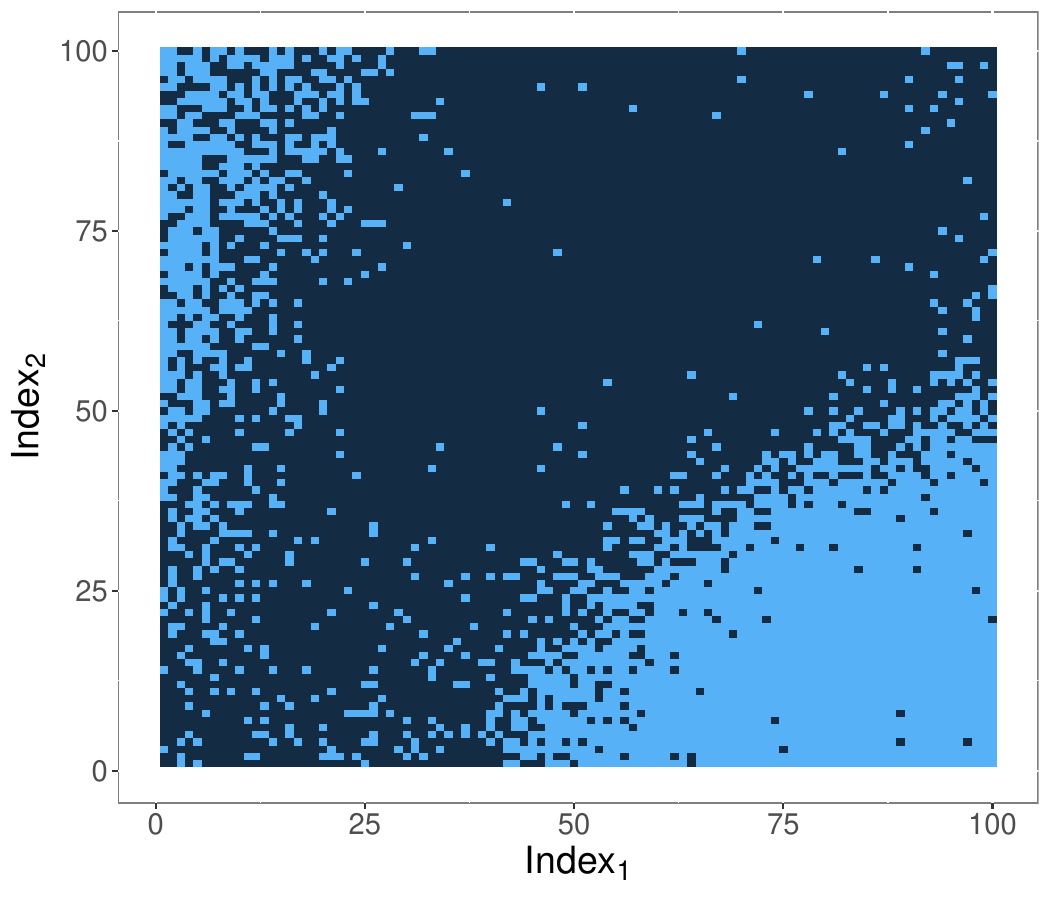} & \includegraphics[scale=0.35]{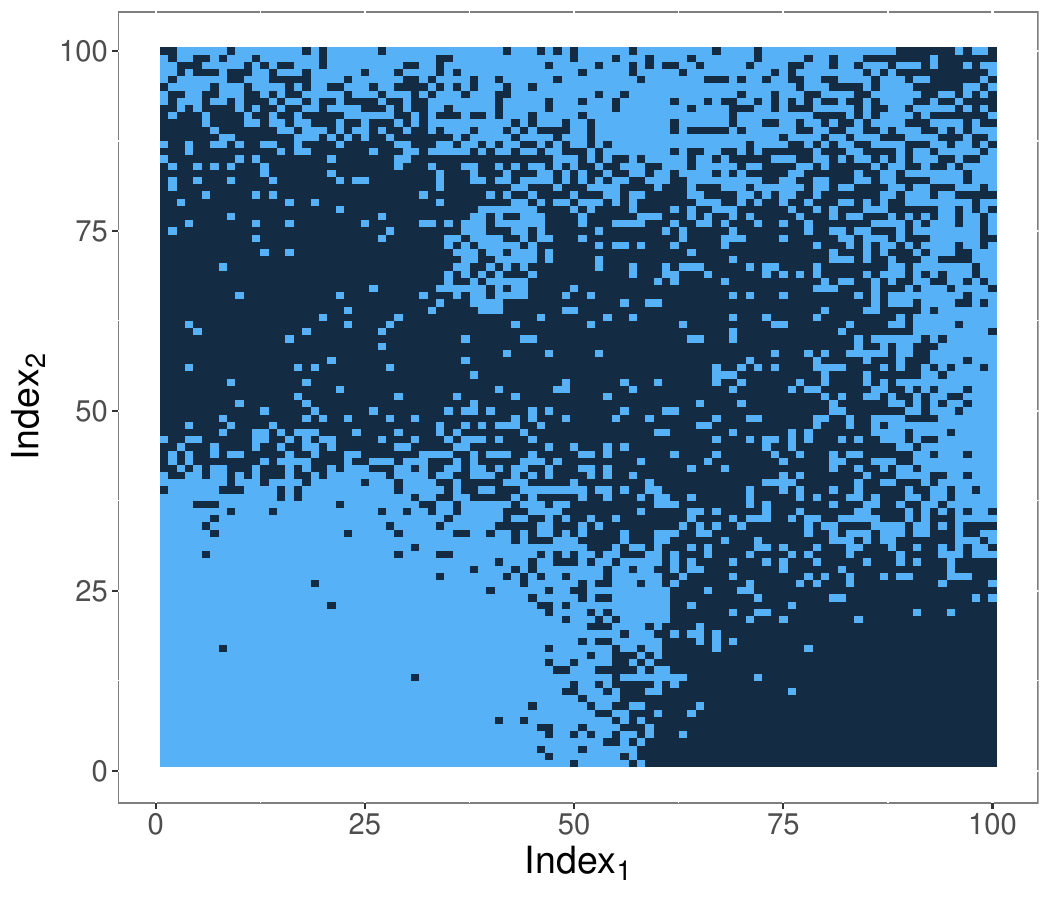}
\end{tabular}
\caption{Examples of simulation lesion status with low to high power heterogeneity (top to bottom) for 500 true signals out of $M=10000$ tests. Light blue refers to lesion.}\label{fig.lesionstatus.example}
\end{figure}

In Figure \ref{fig.bin.simfdr2} and Figure \ref{fig.bin.simpower2}, we present additional summarized results of simulation studies with more parameters settings for WABH (WABH procedures with CAMT estimated non-null probability and MMW $\eta$ where $\tau = 0.5$ or $0.9$, WABH procedure with AdaPT estimated non-null probability and MMW $\eta$ where $\tau = 0.9$, WABH procedure with Storey constant non-null probability). All procedures have acceptable $FDR$ values, and the WABH procedures with AdaPT or CAMT estimated non-null probability have $FDR$ values which are near 0.05. The WABH procedure with constant non-null probability have similar conservative $FDR$ values to the Regular BH, Adaptive BH and Ten Rule procedures. WABH procedures with CAMT estimated non-null probability have relatively larger power than the other WABH procedures. When the spatial dependence is large, the WABH procedure with AdaPT estimated non-null probability has larger power than Regular BH, but less power than the WABH with CAMT estimated non-null probability. The power of WABH procedure with constant non-null probability has not much difference with the Regular BH procedure. In Figure \ref{fig.signal.example}, we can find that the signals will change from randomly spread to clustered when the spatial dependence increases. Figure \ref{fig.lesionstatus.example} shows the lesion status examples for low, moderate and high power heterogeneity.

\label{lastpage}

\end{document}